\documentclass[pdflatex,sn-mathphys-num]{sn-jnl}%

\usepackage{graphicx}%
\usepackage{multirow}%
\usepackage{booktabs}%

\usepackage{amsmath,amssymb,amsfonts}%
\usepackage{amsthm}%
\usepackage{mathrsfs}%

\usepackage{algorithm}%
\usepackage{algorithmicx}%
\usepackage{algpseudocode}%
\usepackage{listings}%

\usepackage{float}
\usepackage{placeins}
\usepackage{wrapfig}
\usepackage{sidecap}
\usepackage[aboveskip=1pt,labelfont=bf,labelsep=period,singlelinecheck=off]{caption}
\usepackage{enumitem}
\usepackage{comment}

\usepackage{tikz}
\usepackage{quantikz}
\usetikzlibrary{positioning, shapes.geometric, calc, arrows.meta}

\usepackage[utf8]{inputenc}
\usepackage[T1]{fontenc}
\usepackage{lmodern}   % scalable Latin Modern fonts
\usepackage{microtype} % load after the font packages
\DisableLigatures[f]{encoding = *, family = * }

\usepackage{xcolor}%
\definecolor{Gray}{gray}{.25} % for figure captions (or other subtle emphasis)

\usepackage{nameref}
\usepackage{hyperref}

\usepackage[title]{appendix}%
\usepackage{textcomp}%
\usepackage{manyfoot}%

\usepackage{changepage} % adjustwidth environment

\usepackage{lastpage,fancyhdr}
\usepackage{epstopdf}
\makeatletter
\renewcommand{\@biblabel}[1]{\quad#1.}
\makeatother

\theoremstyle{thmstyleone}%
\theoremstyle{thmstyletwo}%

\theoremstyle{thmstylethree}%
\begin{document}

\title[Classical $\mathrm{SU}(2)$ Models Match or Exceed Shallow Variational Quantum Circuits on Vision Benchmarks]
{Classical $\mathrm{SU}(2)$ Models Match or Exceed Shallow Variational Quantum Circuits on Vision Benchmarks}

	\author[1]{\fnm{Christopher P.} \sur{Fulton}}
	\author[2]{\fnm{Irene} \sur{Tsapara}}
	\author*[3]{\fnm{Lawrence V.} \sur{Fulton}}\email{fultonl@bc.edu}
	\equalcont{These authors contributed equally to this work.}
	
	\affil[1]{%
		\orgname{United States Air Force Test Pilot School},
		\orgaddress{\city{Edwards Air Force Base}, \state{California}, \postcode{93524}, \country{USA}}
	}
	
	\affil[2]{%
		\orgname{Department of Engineering, Data, and Computer Science; National University},
		\orgaddress{\city{San Diego}, \state{California}, \postcode{92123}, \country{USA}}
	}
	
	\affil*[3]{%
		\orgname{Applied Analytics, Boston College},
		\orgaddress{\street{140 Commonwealth Ave}, \city{Chestnut Hill}, \state{MA}, \postcode{02467}, \country{USA}}
	}

	
\abstract{Quaternion-valued neural networks and variational quantum circuits (VQCs) both derive local transformations from $\mathrm{SU}(2)$ geometry, yet their relative performance on classical supervised learning tasks remains poorly understood. We present a controlled comparison in which real-valued, quaternion-valued, and quantum classification heads operate on identical frozen feature representations across MNIST, FashionMNIST, and CIFAR-10. CIFAR-10 experiments include both a learned 16-dimensional bottleneck and frozen ImageNet-pretrained ResNet18 features to separate architectural effects from representation quality. Quaternion classifiers consistently match or closely approach real-valued baselines while substantially outperforming shallow VQCs. On MNIST and FashionMNIST, quaternion networks achieve near-equivalence with real-valued multilayer perceptrons, whereas product-state VQCs exhibit lower accuracy and substantially higher computational cost. On CIFAR-10, quaternion networks retain 94--97\% of real-valued performance across both feature regimes and remain stable under a 32-fold increase in feature dimensionality. Product-state quantum circuits underperform quaternion classifiers across all benchmarks, while entanglement provides only modest gains on grayscale datasets and reverses under pretrained CNN features, corresponding to a 9.25 percentage-point degradation relative to the product-state circuit. Optimization diagnostics show that Fubini--Study / quantum Fisher information natural-gradient methods improve geometric alignment but do not materially improve short-horizon loss reduction relative to Adam. A Friedman test on the five-seed MNIST evaluation provides evidence of a non-random model ordering ($\chi^2 = 12.796$, $p = 0.0051$, $n = 5$), with post-hoc Wilcoxon signed-rank tests yielding large effect sizes ($d > 5$) for all QuatNet vs.\ quantum comparisons on MNIST. For FashionMNIST and CIFAR-10, large effect sizes ($d > 2.0$) serve as the primary inferential statistic given $n = 3$. These results indicate that, on classical vision benchmarks lacking intrinsic quantum structure, quaternion networks provide efficient and stable $\mathrm{SU}(2)$ alternatives to shallow variational quantum circuits. The findings suggest that shared local $\mathrm{SU}(2)$ geometry and shallow entanglement are insufficient, within the shallow circuit regime studied here, to confer practical quantum advantage on classical image-classification tasks. The conclusions are bounded to shallow, measurement-limited variational circuits operating on classical image-classification tasks without intrinsic quantum structure.}
	
\keywords{Quaternion neural networks, Variational quantum circuits, SU(2) geometry, Quantum machine learning}
	
\maketitle  

%%%%%%%% INTRODUCTION %%%%%%%%%%

\section*{Introduction}

Quaternion-valued neural networks provide compact and expressive architectures for modeling structured correlations in visual, auditory, and spatial signals \cite{parcollet2019survey,parcollet2019quaternion,gaudet2018deep}. Hypercomplex representations based on the complex numbers $\mathbb{C}$ and quaternions $\mathbb{H}$ have long been used in signal-processing applications \cite{mandic2009complex}, where algebraic structure efficiently encodes phase relationships and spatial couplings. Unit quaternions parameterize three-dimensional rotations and form a Lie group isomorphic to $\mathrm{SU}(2)$ \cite{kuipers1999quaternions}, yielding smooth constrained parameterizations well suited to gradient-based optimization. These properties have enabled quaternion networks to achieve competitive performance under reduced parameter budgets across a range of signal-processing and vision tasks \cite{parcollet2019quaternion,gaudet2018deep}.

In parallel, variational quantum circuits (VQCs) have emerged as a prominent paradigm in quantum machine learning. VQCs are constructed from trainable unitary operations, with single-qubit gates drawn from $\mathrm{SU}(2)$ and combined through tensor products and entangling operations \cite{schuld2020circuit,beer2020training}. Their expressive power is often attributed to both local rotation dynamics and multi-qubit entanglement \cite{sim2019expressibility}. A central motivation for VQCs is the possibility of quantum-enhanced feature spaces induced by quantum evolution \cite{havlicek2019supervised}. Recent advances in GPU-accelerated simulation and adjoint differentiation have enabled systematic empirical evaluation of these models at practical scales \cite{bergholm2018pennylane,nvidia2021cuquantum}.

Despite these parallel developments, the relationship between quaternion networks and VQCs remains largely unexplored, even though both rely on trainable transformations derived from $\mathrm{SU}(2)$. Quaternion layers implement normalized Hamilton products with unit quaternions, whereas VQCs implement unitary evolution through parameterized quantum gates. Although these models differ fundamentally in state space, measurement, and entanglement structure, their shared local geometry raises a central question: to what extent can classical $\mathrm{SU}(2)$ structure reproduce the learning behavior of shallow quantum circuits on classical data?

To address this question, we construct a controlled comparison framework in which real-valued, quaternion-valued, and quantum classification heads operate on identical frozen feature representations. Experiments are conducted on MNIST \cite{lecun1998mnist}, FashionMNIST \cite{xiao2017fashion}, and CIFAR-10 \cite{krizhevsky2009learning} under both low-dimensional learned bottlenecks and frozen ImageNet-pretrained ResNet18 features. By holding feature representations fixed within each regime, this design isolates classification-head geometry from feature extraction. Varying representation strength across regimes then allows direct evaluation of how local $\mathrm{SU}(2)$ structure, entanglement, and measurement interact under different feature qualities. Circuits are intentionally restricted to depth 3 to preserve trainability. Increasing depth improves expressivity but introduces barren plateau phenomena and substantial optimization and scaling difficulties~\cite{mcclean2018barren,huembeli2021characterizing}.

Using this framework, we find that quaternion-valued networks consistently match or closely approach real-valued baselines while outperforming the shallow VQCs evaluated here across all datasets and feature regimes. The results support recent perspectives arguing that claims of quantum advantage should be evaluated relative to strong classical baselines and problem structure rather than inferred from model class alone \cite{schuld2022quantum}.

\subsection*{Unit quaternions and their correspondence to $\mathrm{SU}(2)$}

A quaternion $q \in \mathbb{H}$ is written as $q = w + xi + yj + zk$, where $w,x,y,z \in \mathbb{R}$ and the imaginary units satisfy Hamilton's multiplication rules $i^2 = j^2 = k^2 = ijk = -1$. Quaternion multiplication (Hamilton product) is associative but noncommutative. The quaternion conjugate is $q^\ast = w - xi - yj - zk$, and the norm satisfies $\lVert q\rVert^2 = qq^\ast = w^2 + x^2 + y^2 + z^2$. Unit quaternions obey $\lVert q\rVert = 1$ and form the three-sphere $S^3$.

The same manifold appears in quantum computation through the Lie group $\mathrm{SU}(2)$, whose elements act as single-qubit rotations. The Lie group of unit quaternions is isomorphic to $\mathrm{SU}(2)$ via the mapping \cite{kuipers1999quaternions}
\[
\Phi:\mathbb{H} \longrightarrow \mathrm{SU}(2),
\qquad
w + xi + yj + zk \longmapsto
\begin{pmatrix}
	w + iz & y + ix \\
	-y + ix & w - iz
\end{pmatrix}.
\]

For unit quaternions, $\Phi(q)$ is unitary with determinant one, and multiplication is preserved under $\Phi(q_1 q_2)=\Phi(q_1)\Phi(q_2)$. The map $\Phi$ is an isomorphism of Lie groups with trivial kernel, establishing $S^3 \cong \mathrm{SU}(2)$. A separate covering map $\pi: \mathrm{SU}(2) \to \mathrm{SO}(3)$ has kernel $\{\pm I\}$, under which antipodal unit quaternions correspond to the same rotation in $\mathbb{R}^3$, making $\mathrm{SU}(2)$ a double cover of $\mathrm{SO}(3)$.

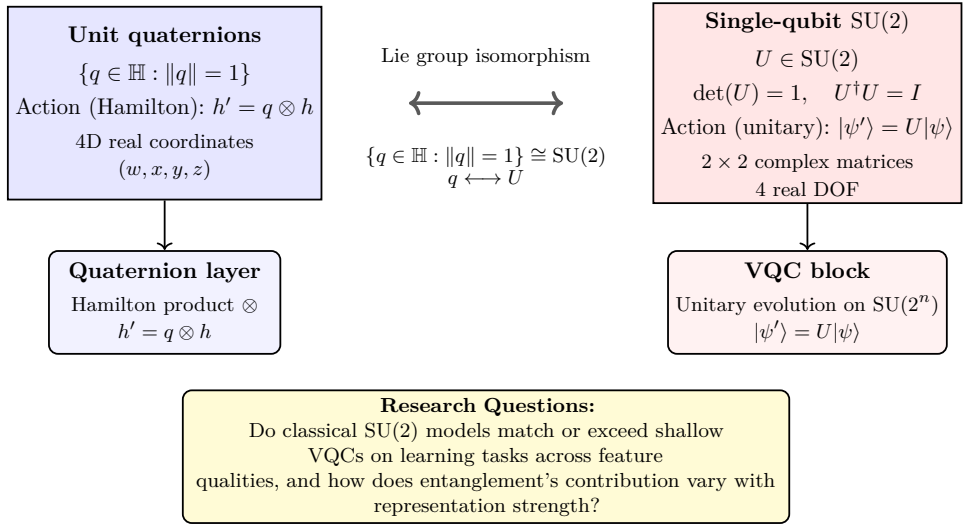
\begin{figure}[H]
	\noindent\makebox[\textwidth][l]{%
		\hspace*{0cm}%
		\scalebox{0.85}{
		\begin{tikzpicture}[scale=1.0]
			
			% Left: Quaternion operations
			\node[
			draw, thick, rectangle,
			minimum width=3.6cm, minimum height=3cm,
			fill=blue!10, align=center
			] (quat) at (0,0) {
				\textbf{Unit quaternions} \\[0.5em]
				$\{q \in \mathbb{H} : \|q\| = 1\}$ \\[0.3em]
				Action (Hamilton): $h' = q \otimes h$ \\[0.3em]
				{\small 4D real coordinates} \\
				{\small $(w,x,y,z)$}
			};
			
			% Right: SU(2) operations
			\node[
			draw, thick, rectangle,
			minimum width=3.6cm, minimum height=3cm,
			fill=red!10, align=center
			] (su2) at (10,0) {
				\textbf{Single-qubit $\mathrm{SU}(2)$} \\[0.5em]
				$U \in \mathrm{SU}(2)$ \\[0.3em]
				$\det(U) = 1,\quad U^\dagger U = I$ \\[0.3em]
				Action (unitary): $|\psi'\rangle = U|\psi\rangle$ \\[0.3em]
				{\small $2\times 2$ complex matrices} \\
				{\small 4 real DOF}
			};
			
			% Middle: Isomorphism
			\draw[<->, ultra thick, black!65] (3.8,0) -- (6.2,0);
			\node[above, align=center, font=\small] at (5,0.45) {Lie group isomorphism};
			\node[below, font=\scriptsize, align=center] at (5,-0.55) {
				$\{q \in \mathbb{H} : \|q\| = 1\} \cong \mathrm{SU}(2)$ \\[0.1em]
				$q \longleftrightarrow U$
			};
			
			% Bottom: Implementation (QNN)
			\node[
			draw, thick, rectangle, rounded corners,
			minimum width=3.6cm, minimum height=1.6cm,
			fill=blue!5, align=center
			] (qnn) at (0,-3.1) {
				\textbf{Quaternion layer} \\[0.25em]
				{\small Hamilton product $\otimes$} \\
				{\small $h' = q \otimes h$}
			};
			
			% Bottom: Implementation (VQC)
			\node[
			draw, thick, rectangle, rounded corners,
			minimum width=3.6cm, minimum height=1.6cm,
			fill=red!5, align=center
			] (vqc) at (10,-3.1) {
				\textbf{VQC block} \\[0.25em]
				{\small Unitary evolution on $\mathrm{SU}(2^n)$} \\
				{\small $|\psi'\rangle = U|\psi\rangle$}
			};
			
			% Connections
			\draw[->, thick] (quat) -- (qnn);
			\draw[->, thick] (su2) -- (vqc);
			
			% Research question
			\node[
			draw, thick, rounded corners,
			fill=yellow!20, align=center,
			text width=9.2cm, font=\small
			] at (5,-5.5) {
				\textbf{Research Questions:} \\
				Do classical $\mathrm{SU}(2)$ models match or exceed shallow \\
				VQCs on learning tasks across feature \\
				qualities, and how does entanglement's contribution vary with \\
				representation strength?
			};
			
		\end{tikzpicture}%
	}
	}%
\caption{Schematic correspondence between unit quaternions and single-qubit $\mathrm{SU}(2)$ operations. Both parameterize transformations on $S^3$, but quaternion layers preserve continuous four-dimensional representations through the Hamilton product, whereas VQCs apply unitary evolution followed by measurement.}
\label{fig:qnn_vqc_comparison}
\end{figure}

This correspondence implies that quaternion layers and single-qubit VQC blocks optimize over the same local rotation group (Fig.~\ref{fig:qnn_vqc_comparison}). The comparison therefore concerns shared parameter geometry rather than computational equivalence.

Several structural differences remain important. Quaternion layers operate on structured representations isomorphic to $\mathbb{R}^{4n}$, whereas quantum circuits evolve states in the $2^n$-dimensional Hilbert space $\mathbb{C}^{2^n}$. Quaternion networks preserve intermediate representations throughout the forward pass, while VQCs ultimately compress quantum states into a small set of measurement expectations. In the four-qubit bottleneck and eight-qubit ResNet18 architectures studied here, a 16- or 256-dimensional complex quantum state is reduced to six or twelve expectation values respectively, discarding amplitude and phase information retained in quaternion layers. Finally, quaternion Hamilton products implement only local $\mathrm{SU}(2)$ rotations, whereas quantum circuits additionally permit entanglement-induced nonseparable correlations.

These distinctions establish that the $\mathrm{SU}(2)$ correspondence is a local structural analogy motivating the comparison, not a claim of computational equivalence. The experiments therefore isolate whether shared local $\mathrm{SU}(2)$ geometry alone is sufficient for performance parity in the shallow regime studied here, and under what conditions entanglement provides measurable benefit.

\section*{Contributions}

\begin{enumerate}
	\item \textbf{Quaternion-valued $\mathrm{SU}(2)$ models closely match real-valued baselines while outperforming shallow VQCs.} Across MNIST, FashionMNIST, and CIFAR-10 under learned bottleneck and pretrained CNN-feature settings, quaternion classifiers match or closely approach real-valued performance while consistently exceeding the shallow quantum classifiers evaluated here. On CIFAR-10, quaternion classifiers retain 94--97\% of real-valued performance across the learned bottleneck and pretrained ResNet18 regimes.
	
	\item \textbf{Shallow VQCs exhibit persistent degradation despite shared local $\mathrm{SU}(2)$ structure.} Under matched frozen-feature conditions, the shallow quantum classifiers evaluated here generally underperform quaternion models while incurring substantially greater computational cost, variance, and optimization sensitivity. The largest degradation occurs for the entangled quantum model under frozen ResNet18 features, suggesting that shared local $\mathrm{SU}(2)$ geometry and shallow ring entanglement are not sufficient, by themselves, to confer practical advantage on the classical vision tasks studied here.
	
	\item \textbf{The effect of entanglement reverses as representation quality increases.} Entanglement provides modest gains on simple grayscale datasets, no measurable benefit on bottleneck CIFAR-10, and substantial degradation under pretrained ResNet18 features, suggesting an interaction between shallow entangling dynamics and rich classical representations.
\end{enumerate}

These results provide a framework for disentangling geometric structure from genuinely quantum contributions in near-term quantum machine learning.

%%%%%%%% METHODS %%%%%%%%%%

\section*{Materials and Methods}

\begin{figure}[!ht]
	\begin{adjustwidth}{-2.2cm}{0cm}
		\centering
		\begin{tikzpicture}[
			scale=0.95,
			every node/.style={font=\small},
			box/.style={rectangle, draw, thick, rounded corners=3pt, 
				minimum width=2.6cm, minimum height=1.3cm, align=center},
			arrow/.style={-Latex, line width=1pt}
			]
			
			% Define colors
			\definecolor{datacol}{RGB}{51,102,204}
			\definecolor{featcol}{RGB}{220,120,0}
			\definecolor{headcol}{RGB}{0,153,76}
			\definecolor{traincol}{RGB}{153,51,153}
			\definecolor{evalcol}{RGB}{204,0,0}
			
			% Nodes
			\node[box, fill=datacol!15, draw=datacol!60] (data) at (0,0) 
			{\textbf{Datasets}\\[2pt]\footnotesize MNIST, Fashion,\\CIFAR-10};
			
			\node[box, fill=featcol!15, draw=featcol!60, right=.5cm of data] (features) 
			{\textbf{Frozen Features}\\[2pt]\footnotesize 16-D bottleneck\\or 512-D ResNet18};
			
			\node[box, fill=headcol!15, draw=headcol!60, right=.5cm of features] (heads) 
			{\textbf{Classification}\\[2pt]\footnotesize RealNet, QuatNet\\Quantum variants};
			
			\node[box, fill=traincol!15, draw=traincol!60, right=.5cm of heads] (opt) 
			{\textbf{Training}\\[2pt]\footnotesize Adam optimizer\\3 random seeds};
			
			\node[box, fill=evalcol!15, draw=evalcol!60, right=.5cm of opt] (eval) 
			{\textbf{Evaluation}\\[2pt]\footnotesize Accuracy\\Time};
			
			% Arrows
			\draw[arrow, black!70] (data) -- (features);
			\draw[arrow, black!70] (features) -- (heads);
			\draw[arrow, black!70] (heads) -- (opt);
			\draw[arrow, black!70] (opt) -- (eval);

			\node[font=\scriptsize\bfseries] at ($(data)!0.5!(eval) + (0,-1.1cm)$) 
			{Experimental Controls};
			
			\node[font=\scriptsize, align=center] at ($(data)!0.5!(eval) + (0,-1.7cm)$)
			{Shared stratified subsets (1,500 train + 300 test per class)\\
				Fixed seeds \{42, 123, 456\} -- Matched data exposure};
			
		\end{tikzpicture}
	\end{adjustwidth}
\noindent\hspace*{-2.2cm}\begin{minipage}{\dimexpr\textwidth+2.5cm\relax}
\caption{Experimental pipeline overview. Models are compared under matched data, feature, and training conditions across three datasets and two feature regimes: a 16-dimensional learned bottleneck and frozen 512-dimensional ResNet18 embeddings.}
\label{fig:methods_roadmap}
\end{minipage}
\end{figure}
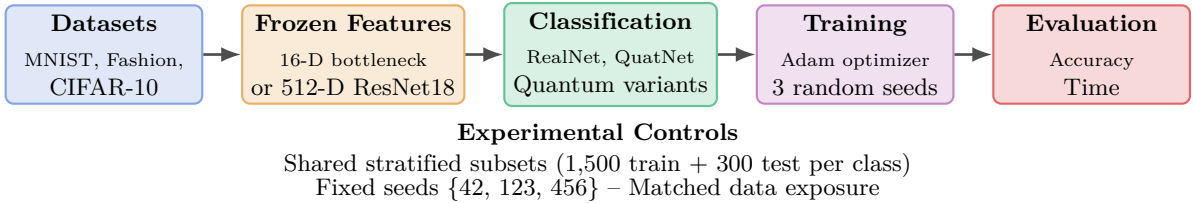

The experiments are designed to evaluate representational geometry in classical and quantum-inspired classifiers under tightly controlled conditions. Data exposure, feature representations, optimization settings, and random seed initialization are held fixed across models to enable controlled head-to-head comparisons. Performance differences can therefore be attributed primarily to the inductive bias and parameterization of the classification heads rather than to dataset imbalance, representation quality, or training variation. The experimental pipeline (Fig.~\ref{fig:methods_roadmap}) proceeds through datasets and preprocessing, frozen feature extraction, classification heads, optimization and training, and diagnostic evaluation.

\subsection*{Datasets and preprocessing}

We evaluate three standard image-classification benchmarks: MNIST handwritten digits \cite{lecun1998mnist}, FashionMNIST clothing items \cite{xiao2017fashion}, and CIFAR-10 natural images \cite{krizhevsky2009learning}. These datasets provide a progression from low-dimensional grayscale structure (MNIST), to visually richer grayscale imagery with greater intra-class variability (FashionMNIST), to full-color natural images with substantially higher semantic complexity (CIFAR-10).

MNIST and FashionMNIST consist of $28\times28$ grayscale images drawn from ten balanced classes. CIFAR-10 consists of $32\times32$ RGB images from ten object categories with substantial background variation and semantic diversity.

All datasets are normalized by subtracting the mean and dividing by the standard deviation. MNIST and FashionMNIST use a global scalar mean and standard deviation computed from the training set. CIFAR-10 uses per-channel normalization, with training-set statistics in the learned-bottleneck regime and standard ImageNet statistics in the pretrained ResNet18 regime.

To enforce strict class balance, we construct stratified subsets by randomly selecting $1{,}500$ training and $300$ test examples per class (seed $42$), yielding $15{,}000$ training and $3{,}000$ test images per dataset, except in the targeted sample-scaling scenario evaluations where training set size is deliberately varied. Real-valued and quaternion models use batch sizes of $128$ for training and $256$ for evaluation.

\subsection*{Frozen feature extraction}

To isolate classification-head effects, feature representations are frozen across all models within each regime, preventing convolutional feature learning from confounding comparisons between head architectures. We examine two feature-extraction regimes. The first uses a deliberately low-capacity learned bottleneck to maximize sensitivity to head-level geometry. The second replaces this bottleneck with a richer pretrained convolutional representation to test whether observed relationships persist under stronger classical features.

In the learned-bottleneck regime used for MNIST, FashionMNIST, and CIFAR-10, inputs are mapped through a shared feature extractor $f_\theta : \mathbb{R}^d \to \mathbb{R}^{16}$ defined by $f_\theta(x)=\tanh(Wx+b)$, where $d=784$ for grayscale images and $d=3072$ for RGB images. The extractor consists of a single linear layer followed by a bounded $\tanh$ activation, which stabilizes quaternion normalization and quantum angle encoding. The extractor is trained once as part of the real-valued baseline (seed 42) and then frozen. The resulting feature vector $h=f_\theta(x)$ is used as fixed input for all quaternion and quantum heads.

For CIFAR-10, we additionally evaluate a frozen ImageNet-pretrained ResNet18 backbone (torchvision default weights), using 512-dimensional embeddings $\phi(x)\in\mathbb{R}^{512}$ extracted from the final average pooling layer. Only the classification heads are optimized, allowing direct comparison between low-dimensional learned bottlenecks and high-capacity pretrained representations.

Quantum models are retained in the pretrained-CNN regime despite increased computational cost. The circuit is scaled to eight qubits, with the classical feature selector expanded to map 512-dimensional inputs to eight encoding angles. Training time increases from approximately four to eighteen hours per seed, but GPU-accelerated simulation remains computationally feasible. Across both regimes, frozen feature extraction ensures that observed performance differences arise from the classification heads rather than unequal preprocessing or end-to-end co-adaptation.

\subsection*{Real-valued baseline}

The real-valued classifier (RealNet) serves as the reference architecture against which quaternion and quantum classifiers are evaluated. In the learned-bottleneck regime, RealNet is implemented as a two-layer multilayer perceptron operating on frozen 16-dimensional features, mapping $\mathbb{R}^{16}\rightarrow\mathbb{R}^{64}\rightarrow\mathbb{R}^{10}$ with a ReLU activation after the hidden layer. The head contains 1{,}738 trainable parameters, excluding the frozen 12{,}560-parameter feature extractor.

For the pretrained ResNet18 control experiment, RealNet operates on frozen 512-dimensional embeddings using the same two-layer structure with an expanded hidden layer, mapping $\mathbb{R}^{512}\rightarrow\mathbb{R}^{128}\rightarrow\mathbb{R}^{10}$. The classification head contains 66{,}954 trainable parameters, excluding the frozen 11.2M-parameter ResNet18 backbone.

\subsection*{Quaternion classifier}

The quaternion classifier (QuatNet) replaces real-valued operations with quaternion-valued layers while preserving identical frozen feature representations across regimes. In the primary experimental setting, the frozen 16-dimensional feature vector is reshaped into four quaternions, $\mathbb{R}^{16}\rightarrow\mathbb{H}^4$, by grouping features into $(w,x,y,z)$ components. The classification head consists of quaternion linear layers mapping $\mathbb{H}^4\rightarrow\mathbb{H}^{16}\rightarrow\mathbb{H}^{10}$, with class logits obtained from the real component of each output quaternion. The quaternion head contains 1{,}000 trainable parameters, representing a 1.7-fold reduction relative to the real-valued baseline.

For the pretrained ResNet18 control experiment, frozen 512-dimensional embeddings are reshaped into 128 quaternions, $\mathbb{R}^{512}\rightarrow\mathbb{H}^{128}$, and processed through quaternion linear layers mapping $\mathbb{H}^{128}\rightarrow\mathbb{H}^{32}\rightarrow\mathbb{H}^{10}$. This configuration contains 17{,}832 trainable parameters, corresponding to a 3.7-fold reduction relative to the real-valued baseline operating on the same features.

Each quaternion layer applies Hamilton products between learned unit-quaternion weights and input quaternions \cite{parcollet2019quaternion,gaudet2018deep},
\[
\mathbf{q}^{(l+1)}_j = b_j + \sum_i W_{ji} \otimes \mathbf{q}^{(l)}_i,
\qquad
W_{ji} \gets \frac{W_{ji}}{\|W_{ji}\|+\varepsilon}.
\]
Weights are initialized as unit quaternions and renormalized after each parameter update. Because weighted sums and bias addition do not preserve unit norm, layer outputs are renormalized prior to the $\tanh$ nonlinearity.

\subsection*{Quantum classifiers}

The quantum classifiers are implemented as a depth-3 VQC with three data re-uploading layers (Fig.~\ref{fig:vqc_circuits}). The circuit operates on four qubits in the learned 16-dimensional bottleneck regime and eight qubits in the pretrained ResNet18 regime. Across regimes and entanglement variants, circuit depth, re-uploading structure, and measurement dimensionality are held fixed, while qubit count and the associated feature-selector mapping ($\mathbb{R}^{16}\!\rightarrow\!\mathbb{R}^{4}$ versus $\mathbb{R}^{512}\!\rightarrow\!\mathbb{R}^{8}$) are adjusted to match input dimensionality.

Measurement dimensionality is intentionally constrained to isolate the effect of local $\mathrm{SU}(2)$ parameter geometry; expanding the measurement space would effectively introduce an additional classical post-processing layer. Deeper circuits were not considered because increased depth introduces additional optimization and scaling confounds, including barren plateaus, simulator cost, noise, and optimizer instability \cite{mcclean2018barren}.

In the learned-bottleneck regime, the frozen 16-dimensional feature vector is mapped to four encoding angles through a trainable feature selector $g_\phi:\mathbb{R}^{16}\rightarrow\mathbb{R}^{4}$ defined by $z=\tanh(W_\phi h+b_\phi)$, where $W_\phi\in\mathbb{R}^{4\times16}$ and $b_\phi\in\mathbb{R}^{4}$. The parameters of $g_\phi$ are optimized jointly with the quantum circuit, while the shared preprocessor $f_\theta$ remains fixed.

For the pretrained ResNet18 regime, the circuit is scaled to eight qubits, with the selector mapping $\mathbb{R}^{512}\rightarrow\mathbb{R}^{8}$ using the same $\tanh(W_\phi h+b_\phi)$ parameterization and $W_\phi\in\mathbb{R}^{8\times512}$. Total trainable parameters increase from 162 to 4{,}282, primarily through expansion of the classical feature selector, while the ResNet18 backbone remains frozen throughout training.

\subsubsection*{Circuit architecture and entanglement variants}

\begin{figure}[h]
	\centering
	\includegraphics[width=\textwidth]{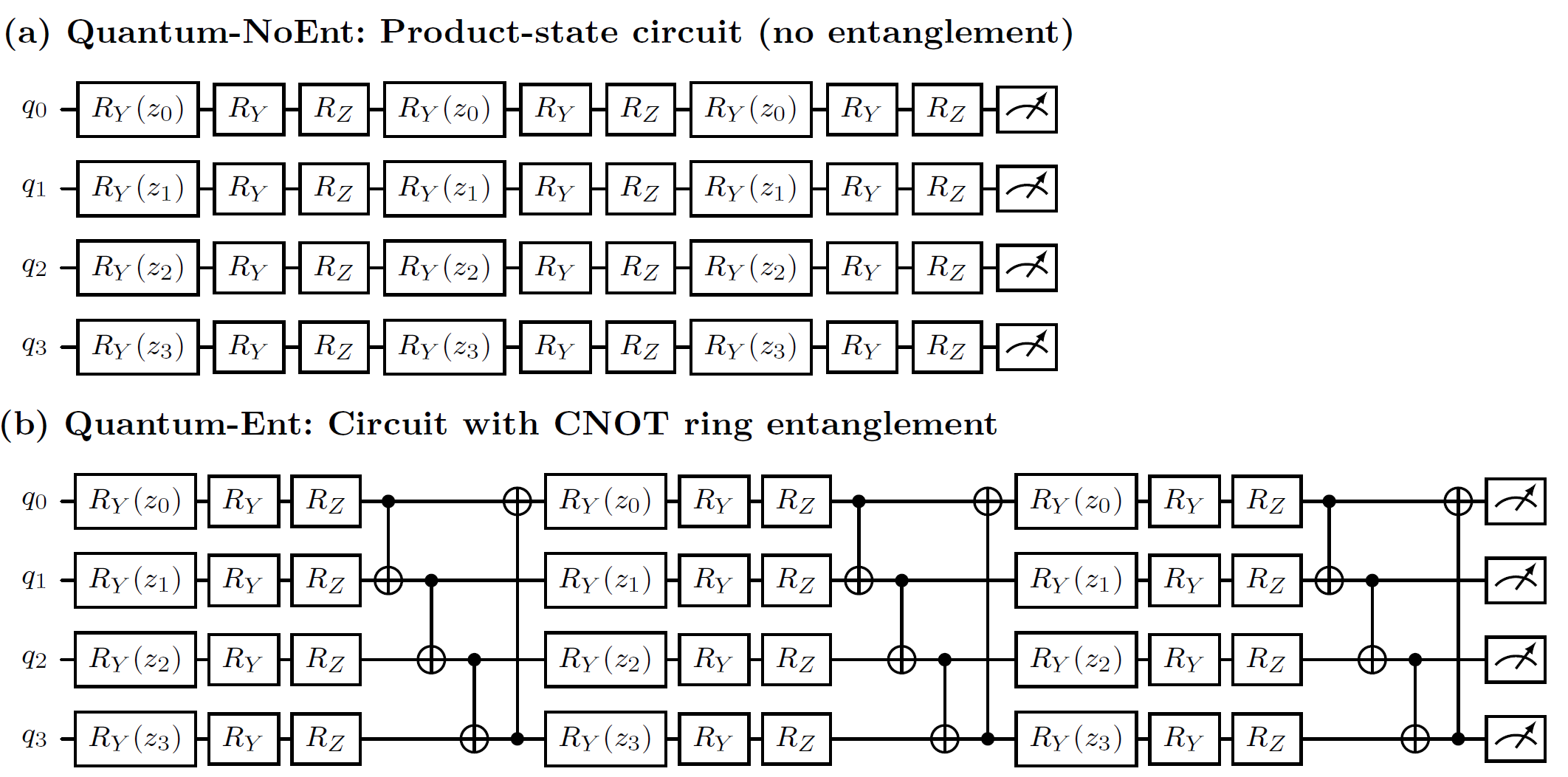}
	\caption{Depth-3 VQC architectures used in the four-qubit MNIST/FashionMNIST regime. Panel (a) shows the product-state model (Quantum-NoEnt), and panel (b) adds a ring of CNOT gates (Quantum-Ent). Each circuit uses data re-uploading with trainable rotations and outputs six observables passed to a linear classifier.}
	\label{fig:vqc_circuits}
\end{figure}

Each circuit layer applies the data re-uploading strategy of Pérez-Salinas et al.~\cite{perezsalinas2020data} through alternating feature encoding and trainable single-qubit rotations. For each layer $\ell\in\{1,2,3\}$ and qubit $i$, the circuit applies $R_Y(z_i)$ followed by trainable rotations $R_Y(\theta^{(\ell)}_{i,Y})$ and $R_Z(\theta^{(\ell)}_{i,Z})$. In the entangled variant (Quantum-Ent), each layer additionally applies a ring of CNOT gates coupling qubits sequentially ($q_0\to q_1\to q_2\to q_3\to q_0$); the product-state variant (Quantum-NoEnt) omits entangling gates (Fig.~\ref{fig:vqc_circuits}). In the pretrained ResNet18 regime, the same depth-3 structure is retained while scaling from four to eight qubits and from 6 to 12 observables.

Measurements are performed at circuit termination. In the four-qubit bottleneck regime, the circuit outputs six observables $(\langle Z_0\rangle,\langle Z_1\rangle,\langle Z_2\rangle,\langle Z_3\rangle,\langle Z_0 Z_1\rangle,\langle Z_2 Z_3\rangle)$ passed to a linear classifier $\mathbb{R}^6\rightarrow\mathbb{R}^{10}$. In the eight-qubit ResNet18 regime, the circuit outputs twelve observables (eight single-qubit Pauli-$Z$ expectations and four $ZZ$ correlations), yielding a linear map $\mathbb{R}^{12}\rightarrow\mathbb{R}^{10}$. In the product-state variant, two-qubit observables factorize and therefore do not increase expressivity; they are retained only to preserve an identical post-processing interface across variants.

Quantum classifiers include a trainable linear feature-selection layer $g_{\phi}$ mapping classical feature vectors to qubit rotation parameters. The same selector architecture is used across all quantum variants, ensuring that differences between Quantum-NoEnt and Quantum-Ent arise from circuit structure rather than feature preprocessing. Although the selector accounts for most trainable parameters in the 512-dimensional regime, it performs only a linear projection into qubit rotation space; nonlinear processing occurs within the variational circuit and measurement.

Because optimization instability is frequently cited as a limitation of variational quantum circuits, we additionally evaluated whether Fubini--Study / quantum Fisher information (FS/QFI, \cite{stokes2020quantum})-aware optimization materially alters training behavior in the shallow regime studied here, providing a direct test of whether optimizer geometry rather than architectural constraints accounts for the observed performance gap (Table~\ref{tab:opt_geometry}). The diagnostic is deliberately short-horizon ($K = 10$ updates) to isolate gradient geometry from downstream representation effects. The goal was not to optimize accuracy, but to determine whether curvature-aware updates produce meaningfully different optimization trajectories than Adam, which would implicate optimizer choice rather than architectural constraints as a primary driver of the observed underperformance.

\begin{table}[h]
	\centering
\caption{Optimization geometry diagnostics in the learned 16-dimensional bottleneck regime, averaged over 30 runs. Losses report final cross-entropy after $K=10$ updates, and cosine similarity measures alignment with the Fubini--Study (FS/QFI) natural-gradient direction.}

	\label{tab:opt_geometry}
	\begin{tabular}{lccccc}
		\toprule
		Optimizer           &   Dataset    & Final loss & Loss SD &     Time (s)     & Cosine vs.\ full FS/QFI \\ \midrule
		Euclidean GD        &    MNIST     &  $2.316$   & $0.039$ & $6.04 \pm 0.36$  &    $0.940 \pm 0.047$    \\
		                    & FashionMNIST &  $2.301$   & $0.052$ & $5.95 \pm 0.41$  &    $0.937 \pm 0.074$    \\
		                    &   CIFAR-10   &  $2.320$   & $0.036$ & $6.09 \pm 0.52$  &    $0.958 \pm 0.031$    \\ \midrule
		Adam                &    MNIST     &  $2.315$   & $0.039$ & $5.97 \pm 0.40$  &    $0.728 \pm 0.073$    \\
		                    & FashionMNIST &  $2.301$   & $0.052$ & $5.92 \pm 0.47$  &    $0.725 \pm 0.067$    \\
		                    &   CIFAR-10   &  $2.320$   & $0.036$ & $5.97 \pm 0.39$  &    $0.689 \pm 0.099$    \\ \midrule
		FS/QFI (diag)       &    MNIST     &  $2.316$   & $0.039$ & $7.33 \pm 0.59$  &    $0.961 \pm 0.030$    \\
		                    & FashionMNIST &  $2.301$   & $0.052$ & $7.32 \pm 0.68$  &    $0.959 \pm 0.041$    \\
		                    &   CIFAR-10   &  $2.320$   & $0.036$ & $7.48 \pm 0.70$  &    $0.974 \pm 0.016$    \\ \midrule
		FS/QFI (block-diag) &    MNIST     &  $2.316$   & $0.039$ & $7.44 \pm 0.62$  &    $0.961 \pm 0.030$    \\
		                    & FashionMNIST &  $2.301$   & $0.052$ & $7.47 \pm 0.72$  &    $0.959 \pm 0.041$    \\
		                    &   CIFAR-10   &  $2.320$   & $0.036$ & $7.37 \pm 0.59$  &    $0.974 \pm 0.016$    \\ \midrule
		FS/QFI (full)       &    MNIST     &  $2.315$   & $0.039$ & $18.11 \pm 1.16$ &         $1.000$         \\
		                    & FashionMNIST &  $2.301$   & $0.052$ & $18.46 \pm 1.51$ &         $1.000$         \\
		                    &   CIFAR-10   &  $2.320$   & $0.036$ & $18.23 \pm 1.36$ &         $1.000$         \\ \bottomrule
	\end{tabular}
\end{table}

Across datasets, FS/QFI diagonal and block-diagonal approximations produced update directions nearly collinear with the full natural-gradient step (cosine similarity $\approx0.96$--$0.97$), whereas Adam shows lower alignment ($\approx0.69$--$0.73$) due to adaptive gradient rescaling. Despite these geometric differences, all optimizers achieve nearly identical short-horizon loss reduction, with final cross-entropy means differing by less than $10^{-3}$. Thus, although Adam deviates substantially from the natural-gradient direction, optimization behavior remains statistically indistinguishable in this regime.

Figure~\ref{fig:architecture} illustrates the complete architecture for all model variants.

\subsection*{Training protocol and evaluation}

All architectures (RealNet, QuatNet, Quantum-NoEnt, and Quantum-Ent) are trained across three random seeds $\{42,123,456\}$, with Python, NumPy, PyTorch, CUDA, and PennyLane seeds fixed and PyTorch deterministic mode enabled. Reported results are mean $\pm$ standard deviation across seeds.

All models minimize cross-entropy loss using Adam with learning rate $10^{-3}$ on fixed stratified training sets. Real-valued and quaternion models are trained for up to 100 epochs with early stopping based on validation accuracy (patience 10). Quantum models use the same criterion but allow up to 200 epochs to accommodate slower variational-circuit convergence. Increasing patience to 20 epochs produced no material change in accuracy.

Quaternion layers are implemented using standard differentiable PyTorch operations with an extrinsic optimize-then-project strategy \cite{absil2008optimization}, in which parameters are updated in Euclidean space and subsequently renormalized to unit norm. This preserves valid unit quaternions without explicit manifold-aware gradient computation.

In the learned-bottleneck regime, the feature extractor $f_\theta$ is trained once using seed~42 and then frozen. In the pretrained CNN regime, the ImageNet-pretrained ResNet18 backbone remains fixed throughout training. For each model, we report test accuracy at the best validation epoch, epochs to early stopping, trainable parameters, and total wall-clock training time, enabling direct comparison of optimization stability and computational efficiency across architectures.

For the primary MNIST benchmark, an extended five-seed evaluation $\{42, 123, 456, 789, 999\}$ was conducted for all four model variants to support effect size estimation and formal statistical significance testing ($\alpha = 0.10$). To assess stability under increased data exposure, targeted sample-scaling evaluations were additionally conducted for QuatNet and Quantum-Ent at 1{,}500, 2{,}000, and 2{,}500 training samples per class under frozen ResNet18 features, and for Quantum-NoEnt at 2{,}000 training samples per class on CIFAR-10 with the learned bottleneck. Full results are reported in the Robustness across seeds and sample sizes subsection.

% ============================================================================
% Architecture Comparison Figure (TikZ)
% ============================================================================

\begin{figure}[!ht]
	\noindent\makebox[\textwidth][l]{%
		\hspace*{-.5cm}%
		\scalebox{.9}{
		\begin{tikzpicture}[
			scale=0.78,
			block/.style={
				rectangle, draw, thick,
				minimum width=2cm, minimum height=0.75cm,
				align=center, font=\small
			},
			frozen/.style={block, fill=gray!20},
			real/.style={block, fill=blue!15},
			quat/.style={block, fill=green!15},
			quantum/.style={block, fill=red!15},
			arrow/.style={->, thick},
			]
			
			% -----------------------
			% Input
			% -----------------------
			\node[block, fill=yellow!20, minimum width=2.3cm] (input) at (0, 0) {
				\textbf{Input Image} \\ (varies by dataset)
			};
			
			% -----------------------
			% Shared Preprocessor
			% -----------------------
			\node[frozen, minimum width=3.2cm, minimum height=1.4cm] (prep) at (0, -2.5) {
				\textbf{Shared Preprocessor} \\
				{\scriptsize (Frozen)} \\
				{\scriptsize 16-D bottleneck or} \\
				{\scriptsize 512-D ResNet18}
			};
			
			\draw[arrow] (input) -- (prep);
			
			% -----------------------
			% Bottleneck
			% -----------------------
			\node[
			draw, dashed, thick, rounded corners,
			minimum width=3.7cm, minimum height=0.65cm,
			fill=orange!10, font=\small
			] (bottleneck) at (0, -4.5) {
				\textbf{Fixed Features}
			};
			
			\draw[arrow] (prep) -- (bottleneck);
			
			% -----------------------
			% Split point
			% -----------------------
			\coordinate (split) at (0, -6.3);
			
			% -----------------------
			% Head Titles (above split)
			% -----------------------
			\node[font=\bfseries, text=blue!70!black]  at (-6.2, -5.7) {Real MLP};
			\node[font=\bfseries, text=green!70!black] at (-1.8, -5.7) {Quaternion};
			\node[font=\bfseries, text=red!70!black]   at ( 3.2, -5.7) {Quantum (No Ent)};
			\node[font=\bfseries, text=red!70!black]   at ( 8.5, -5.7) {Quantum (Ent)};
			
			% -----------------------
			% Real Head
			% -----------------------
			\begin{scope}[xshift=-6.2cm]
				\node[real, minimum height=0.90cm] (real1) at (0, -8.2) {
					Real Linear \\ (varies by regime)
				};
				\node[real, minimum height=0.90cm] (real2) at (0, -10.6) {
					Real Linear \\ (varies by regime)
				};
				\node[block, fill=blue!5, minimum height=0.70cm] (real_out) at (0, -12.7) {
					Logits (10)
				};
				
				\draw[arrow, blue!70] (split) -| (real1);
				\draw[arrow] (real1) -- (real2);
				\draw[arrow] (real2) -- (real_out);
				
				\node[font=\scriptsize] at (0, -13.5) {1.7K (16-D) or};
				\node[font=\scriptsize] at (0, -14) {67K (512-D)};
			\end{scope}
			
			% -----------------------
			% Quaternion Head
			% -----------------------
			\begin{scope}[xshift=-1.8cm]
				\node[quat, minimum height=0.90cm] (quat1) at (0, -8.2) {
					Quat Linear \\ (varies by regime)
				};
				\node[quat, minimum height=0.90cm] (quat2) at (0, -10.6) {
					Quat Linear \\ (varies by regime)
				};
				\node[block, fill=green!5, minimum height=0.70cm] (quat_out) at (0, -12.7) {
					Extract Real \\ Logits (10)
				};
				
				\draw[arrow, green!70!black] (split) -| (quat1);
				\draw[arrow] (quat1) -- (quat2);
				\draw[arrow] (quat2) -- (quat_out);
				
				\node[font=\scriptsize] at (0, -13.8) {1K (16-D) or};
				\node[font=\scriptsize] at (0, -14.3) {18K (512-D)};
			\end{scope}
			
			% -----------------------
			% Quantum Head (No Entanglement)
			% -----------------------
			\begin{scope}[xshift=3.2cm]
				\node[quantum, minimum height=0.70cm, minimum width=2.35cm] (q_select) at (0, -8.2) {
					Feature Selector $g_\phi$\\ (varies by regime)
				};
				\node[quantum, minimum width=2.60cm, minimum height=1.55cm] (q_circuit) at (0, -10.8) {
					\textbf{VQC} (4q/8q, depth-3) \\
					{\scriptsize $\mathrm{SU}(2)$ rotations}
				};
				\node[quantum, minimum height=0.80cm, minimum width=2.35cm] (q_measure) at (0, -13.3) {
					Measure \\ 6/12 expvals
				};
				\node[block, fill=red!5, minimum height=0.70cm, minimum width=2.15cm] (q_out) at (0, -15.2) {
					Linear 6/12 $\to$ 10
				};
				
				\draw[arrow, red!70] (split) -| (q_select);
				\draw[arrow] (q_select) -- (q_circuit);
				\draw[arrow] (q_circuit) -- (q_measure);
				\draw[arrow] (q_measure) -- (q_out);
				
				\node[font=\scriptsize] at (0, -16.1) {162 (16-D) or 4K (512-D)};
			\end{scope}
			
			% -----------------------
			% Quantum Head (With Entanglement)
			% -----------------------
			\begin{scope}[xshift=8.5cm]
				\node[quantum, minimum height=0.70cm, minimum width=2.35cm] (qe_select) at (0, -8.2) {
					Feature Selector $g_\phi$\\ (varies by regime)
				};
				\node[quantum, minimum width=2.60cm, minimum height=1.55cm] (qe_circuit) at (0, -10.8) {
					\textbf{VQC} (4q/8q, depth-3) \\
					{\scriptsize $\mathrm{SU}(2)$ rotations}
				};
				\node[quantum, minimum height=0.80cm, minimum width=2.35cm] (qe_measure) at (0, -13.3) {
					Measure \\ 6/12 expvals
				};
				\node[block, fill=red!5, minimum height=0.70cm, minimum width=2.15cm] (qe_out) at (0, -15.2) {
					Linear 6/12 $\to$ 10
				};
				
				\draw[arrow, red!70] (split) -| (qe_select);
				\draw[arrow] (qe_select) -- (qe_circuit);
				\draw[arrow] (qe_circuit) -- (qe_measure);
				\draw[arrow] (qe_measure) -- (qe_out);
				
				\node[font=\scriptsize] at (0, -16.1) {162 (16-D) or 4K (512-D)};
			\end{scope}
			
			% -----------------------
			% Legend / Notes
			% -----------------------
			\node[
			draw, thick, rounded corners,
			fill=gray!10, text width=14.5cm,
			font=\scriptsize, align=left, inner sep=7pt
			] at (1.5, -18.4) {
				\textbf{Experimental Design:} \\
				\textbullet\ Two feature regimes: 16-D bottleneck (MNIST, FM, CIFAR-10) and 512-D ResNet18 (CIFAR-10) \\
				\textbullet\ Preprocessor frozen after Real baseline training (16-D) or pretrained (ResNet18) \\
				\textbullet\ All heads operate on identical fixed features within each regime \\
				\textbullet\ Isolates the inductive bias of the head architecture, not feature extraction
			};
			
		\end{tikzpicture}%
	}
	}%
	
\caption{Architecture comparison of real-valued, quaternion, and quantum classifiers under shared frozen features. Quaternion heads apply Hamilton products with unit quaternions, while quantum heads use depth-3 variational circuits with and without ring-topology CNOT entanglement.}

		\label{fig:architecture}
	
\end{figure}

\section*{Results}

We first evaluate MNIST and FashionMNIST using a shared 16-dimensional learned bottleneck, then extend the analysis to CIFAR-10 under both the learned bottleneck and frozen 512-dimensional ImageNet-pretrained ResNet18 features. Robustness is assessed through three complementary excursions: a five-seed extension $\{42, 123, 456, 789, 999\}$ of the primary MNIST benchmark for all four model variants, a targeted sample-scaling analysis of QuatNet and Quantum-Ent at 1{,}500, 2{,}000, and 2{,}500 training samples per class under frozen ResNet18 features, and a sample-size robustness check for Quantum-NoEnt at 2{,}000 training samples per class on CIFAR-10 with the learned bottleneck. Full robustness results are reported in the Robustness across seeds and sample sizes subsection.

Across all datasets, quaternion classifiers consistently match or approach real-valued baselines while outperforming both quantum variants (Table~\ref{tab:aggregated_results}, Figure~\ref{fig:performance_comparison}). On MNIST and FashionMNIST, QuatNet achieves near-equivalence with RealNet, while product-state quantum models remain below the quaternion classifier by approximately 6.1 percentage points on MNIST and 2.4 percentage points on FashionMNIST despite substantially longer training times. Entanglement yields only modest gains ($\approx0.6$ percentage points) on grayscale benchmarks. On CIFAR-10 with ResNet18 features, quaternion classifiers retain $97.2\%$ of the real-valued baseline, demonstrating robustness across representation strengths. The effect of entanglement reverses across representation regimes, shifting from modest gains on grayscale datasets to a 9.3-percentage-point degradation under high-quality pretrained features. This reversal persists under increased data exposure: Quantum-Ent achieves $39.07\%$ at 2{,}000 samples per class and $38.13\%$ at 2{,}500 samples per class, remaining substantially below both classical baselines and exhibiting non-monotonic scaling, while QuatNet remains stable at $45.81\%$--$46.54\%$ across the same range.

\subsection*{Real-valued baseline (RealNet)}

RealNet combines the shared 16-dimensional bottleneck with a two-layer multilayer perceptron (16 $\to$ 64 $\to$ 10). On MNIST, FashionMNIST, and CIFAR-10, RealNet achieves $93.54\% \pm 0.52\%$, $84.60\% \pm 0.12\%$, and $40.23\% \pm 0.44\%$ test accuracy, respectively, converging in approximately 21--24 epochs with training times of 10--32 seconds per seed. The decline in accuracy from MNIST to CIFAR-10 reflects increasing visual complexity under the shared 16-dimensional bottleneck.

Using frozen 512-dimensional ResNet18 features on CIFAR-10 increases accuracy to $47.13\% \pm 0.70\%$ (+6.9 percentage points), indicating that the bottleneck regime intentionally constrains representation capacity. The pretrained-feature regime converges in $22.3 \pm 4.0$ epochs with training times of $14.2 \pm 2.5$ seconds per seed.

\subsection*{Quaternion classifier (QuatNet)}

QuatNet replaces the real-valued MLP head with quaternion-valued layers applying Hamilton products with learned unit-quaternion weights. Across datasets and feature regimes, QuatNet closely matches RealNet despite substantially fewer parameters.

On MNIST and FashionMNIST, QuatNet achieves $93.64\% \pm 0.15\%$ and $84.47\% \pm 0.05\%$ test accuracy, closely matching the real-valued baseline while using 1{,}000 trainable parameters versus 1{,}738 for RealNet. Convergence occurs in approximately 24--32 epochs with training times of 57--77 seconds per seed.

On CIFAR-10 with the learned 16-dimensional bottleneck, QuatNet achieves $37.92\% \pm 0.33\%$, retaining $94.3\%$ of RealNet performance. Using frozen 512-dimensional ResNet18 features increases accuracy to $45.81\% \pm 0.16\%$, preserving $97.2\%$ of the real-valued baseline while using 17{,}832 parameters versus 66{,}954 for RealNet.

These results indicate that the imposed $\mathrm{SU}(2)$ structure acts primarily as an inductive bias rather than an expansion in representational capacity.

\subsection*{Quantum classifier without entanglement (Quantum-NoEnt)}

On MNIST, FashionMNIST, and CIFAR-10 with the learned 16-dimensional bottleneck, Quantum-NoEnt achieves $87.52\% \pm 1.02\%$, $82.03\% \pm 0.94\%$, and $35.10\% \pm 0.71\%$ test accuracy, respectively, requiring approximately 4--7 GPU-hours per seed. Using frozen 512-dimensional ResNet18 features increases CIFAR-10 accuracy to $41.71\% \pm 0.94\%$, but performance remains below both the real-valued and quaternion classifiers despite roughly 18 GPU-hours of training per seed.

Across datasets and feature regimes, Quantum-NoEnt consistently underperforms both classical baselines while exhibiting substantially higher computational cost and variance. Although the circuit shares the same local $\mathrm{SU}(2)$ rotation structure as quaternion layers, measurement-induced compression, shallow-circuit expressivity limits, and optimization instability constrain performance on these classical vision tasks, consistent with known trainability challenges in variational quantum circuits \cite{mcclean2018barren}.

\subsection*{Quantum classifier with entanglement (Quantum-Ent)}

For MNIST and FashionMNIST, Quantum-Ent achieves $88.16\% \pm 0.62\%$ and $82.62\% \pm 0.52\%$ test accuracy, respectively, providing modest gains of approximately $0.6$--$0.7$ percentage points over the product-state variant at substantially higher computational cost ($\sim10$ GPU-hours per seed). On CIFAR-10 with the learned 16-dimensional bottleneck, performance drops to $34.92\% \pm 0.22\%$, slightly below the non-entangling circuit.

Using frozen 512-dimensional ResNet18 features produces a marked degradation: Quantum-Ent achieves only $32.46\% \pm 3.18\%$ test accuracy, 9.3 percentage points below the product-state variant despite requiring approximately 18 GPU-hours per seed. The effect of entanglement therefore reverses as feature quality increases. Under rich classical representations, shallow entangling circuits exhibit amplified optimization instability and measurement-induced compression, consistent with known trainability challenges in variational quantum circuits \cite{mcclean2018barren}.

Across all benchmarks and feature regimes, Quantum-Ent remains substantially below both real-valued and quaternion baselines while incurring the highest computational cost. Although entanglement introduces nonclassical correlations unavailable to quaternion layers, these correlations do not translate into improved performance in shallow, measurement-limited circuits.

\subsection*{Head-to-head comparison}

\begin{figure}[!h]
	\begin{adjustwidth}{-2.75cm}{0cm}
		
		\centering
		\begin{tikzpicture}[
			scale=.85,
			every node/.style={font=\small}
			]
			
			% --------------------------------------------------
			% Color-blind friendly palette (Okabe--Ito)
			% --------------------------------------------------
			\definecolor{realcolor}{HTML}{0072B2} % blue
			\definecolor{quatcolor}{HTML}{009E73} % bluish green
			\definecolor{qnoentcolor}{HTML}{D55E00} % vermillion
			\definecolor{qentcolor}{HTML}{CC79A7} % reddish purple
			
			% Marker size control (one knob)
			\def\ms{0.62}
			\def\msquat{0.31} % Half size for QuatNet circle
			
			% --------------------------------------------------
			% Mean-marker glyphs (drawn in TikZ; no external images)
			% Each macro takes a color name as #1
			% --------------------------------------------------
			\newcommand{\MeanReLU}[1]{%
				\tikz[baseline=-0.6ex, scale=\ms]{
					\draw[line width=0.7pt, #1] (-0.35,-0.10) -- (-0.05,-0.10) -- (0.20,0.20) -- (0.35,0.20);
				}%
			}
			
			\newcommand{\MeanQuat}[1]{%
				\tikz[baseline=-0.6ex, scale=\msquat]{
					\draw[line width=0.7pt, #1] (0,0) circle (0.30);
					\node[font=\scriptsize, text=#1] at (0.50,0) {$\mathbf{i}$};
					\node[font=\scriptsize, text=#1] at (-0.50,0) {$\mathbf{j}$};
					\node[font=\scriptsize, text=#1] at (0,0.80) {$\mathbf{k}$};
				}%
			}
			
			\newcommand{\MeanBloch}[1]{%
				\tikz[baseline=-0.6ex, scale=\ms]{
					\draw[line width=0.7pt, #1] (0,0) circle (0.30);
					\draw[line width=0.5pt, #1] (-0.30,0) arc (180:360:0.30 and 0.12); % equator
					\draw[line width=0.5pt, #1] (0,-0.30) arc (-90:90:0.12 and 0.30); % meridian
				}%
			}
			
			\newcommand{\MeanBlochE}[1]{%
				\tikz[baseline=-0.6ex, scale=\ms]{
					\draw[line width=0.7pt, #1] (0,0) circle (0.30);
					\draw[line width=0.5pt, #1] (-0.30,0) arc (180:360:0.30 and 0.12);
					\draw[line width=0.5pt, #1] (0,-0.30) arc (-90:90:0.12 and 0.30);
					\node[font=\scriptsize\bfseries, text=#1] at (0,0) {E};
				}%
			}
			
			% --------------------------------------------------
			% Geometry controls (FOUR panels now)
			% --------------------------------------------------
			\def\xmax{18.5}  % Wider to accommodate 4 panels
			\def\ymax{5.71}
			\def\xsep{0.25*\xmax}     % First separator at 25%
			\def\xseptwo{0.50*\xmax}  % Second separator at 50%
			\def\xsepthree{0.75*\xmax} % Third separator at 75%
			
			% X positions as fractions of panel width
			% Panel 1: MNIST (0-25%)
			\def\xR{0.04*\xmax}
			\def\xQ{0.09*\xmax}
			\def\xN{0.14*\xmax}
			\def\xE{0.19*\xmax}
			
			% Panel 2: FashionMNIST (25-50%)
			\def\xRr{0.29*\xmax}
			\def\xQr{0.34*\xmax}
			\def\xNr{0.39*\xmax}
			\def\xEr{0.44*\xmax}
			
			% Panel 3: CIFAR-10 16-D (50-75%)
			\def\xRc{0.54*\xmax}
			\def\xQc{0.59*\xmax}
			\def\xNc{0.64*\xmax}
			\def\xEc{0.69*\xmax}
			
			% Panel 4: CIFAR-10 512-D (75-100%)
			\def\xRd{0.79*\xmax}
			\def\xQd{0.84*\xmax}
			\def\xNd{0.89*\xmax}
			\def\xEd{0.94*\xmax}
			
			% Common styles
			\tikzset{
				axis/.style={->, thick},
				grid/.style={gray!25},
				sep/.style={dashed, gray},
			}
			
			% ==================================================
			% (a) Test Accuracy
			% ==================================================
			\begin{scope}[yshift=0cm]
				
				% Axes
				\draw[axis] (0,0) -- (\xmax,0);
				\draw[axis] (0,0) -- (0,\ymax);
				
				% Rotated y-axis label (centered)
				\node[rotate=90, anchor=south] at (-0.85,0.5*\ymax) {Test Accuracy};
				
				% Grid + y tick labels (adjusted to include 1.00)
				\foreach \y/\lab in {0/0.30,1.43/0.50,2.86/0.70,4.29/0.90,5.71/1.00}{
					\draw[grid] (0,\y) -- (\xmax-0.2,\y);
					\node[left, font=\scriptsize] at (0,\y) {\lab};
				}
				
				% Separators
				\draw[sep] (\xsep,0) -- (\xsep,\ymax);
				\draw[sep] (\xseptwo,0) -- (\xseptwo,\ymax);
				\draw[sep] (\xsepthree,0) -- (\xsepthree,\ymax);
				
				% Dataset labels
				\node[font=\bfseries] at (0.125*\xmax,\ymax+0.35) {MNIST};
				\node[font=\bfseries] at (0.375*\xmax,\ymax+0.35) {FashionMNIST};
				\node[font=\bfseries, align=center] at (0.625*\xmax,\ymax+0.35) {CIFAR-10\\{\scriptsize (16-D)}};
				\node[font=\bfseries, align=center] at (0.875*\xmax,\ymax+0.35) {CIFAR-10\\{\scriptsize (512-D)}};
				
				% Helper function: map accuracy to y coordinate
				% y = (acc - 0.30) / 0.70 * 5.71
				
				% ---- MNIST ----
				\draw[realcolor, very thick] (\xR,{(0.9293-0.30)/0.70*5.71}) -- (\xR,{(0.9420-0.30)/0.70*5.71});
				\node at (\xR,{(0.9354-0.30)/0.70*5.71}) {\MeanReLU{realcolor}};
				
				\draw[quatcolor, very thick] (\xQ,{(0.9347-0.30)/0.70*5.71}) -- (\xQ,{(0.9383-0.30)/0.70*5.71});
				\node at (\xQ,{(0.9364-0.30)/0.70*5.71}) {\MeanQuat{quatcolor}};
				
				\draw[qnoentcolor, very thick] (\xN,{(0.8670-0.30)/0.70*5.71}) -- (\xN,{(0.8897-0.30)/0.70*5.71});
				\node at (\xN,{(0.8752-0.30)/0.70*5.71}) {\MeanBloch{qnoentcolor}};
				
				\draw[qentcolor, very thick] (\xE,{(0.8770-0.30)/0.70*5.71}) -- (\xE,{(0.8903-0.30)/0.70*5.71});
				\node at (\xE,{(0.8816-0.30)/0.70*5.71}) {\MeanBlochE{qentcolor}};
				
				% ---- FashionMNIST ----
				\draw[realcolor, very thick] (\xRr,{(0.8450-0.30)/0.70*5.71}) -- (\xRr,{(0.8477-0.30)/0.70*5.71});
				\node at (\xRr,{(0.8460-0.30)/0.70*5.71}) {\MeanReLU{realcolor}};
				
				\draw[quatcolor, very thick] (\xQr,{(0.8440-0.30)/0.70*5.71}) -- (\xQr,{(0.8450-0.30)/0.70*5.71});
				\node at (\xQr,{(0.8447-0.30)/0.70*5.71}) {\MeanQuat{quatcolor}};
				
				\draw[qnoentcolor, very thick] (\xNr,{(0.8107-0.30)/0.70*5.71}) -- (\xNr,{(0.8330-0.30)/0.70*5.71});
				\node at (\xNr,{(0.8203-0.30)/0.70*5.71}) {\MeanBloch{qnoentcolor}};
				
				\draw[qentcolor, very thick] (\xEr,{(0.8203-0.30)/0.70*5.71}) -- (\xEr,{(0.8330-0.30)/0.70*5.71});
				\node at (\xEr,{(0.8262-0.30)/0.70*5.71}) {\MeanBlochE{qentcolor}};
				
				% ---- CIFAR-10 (16-D) ----
				\draw[realcolor, very thick] (\xRc,{(0.3973-0.30)/0.70*5.71}) -- (\xRc,{(0.4073-0.30)/0.70*5.71});
				\node at (\xRc,{(0.4023-0.30)/0.70*5.71}) {\MeanReLU{realcolor}};
				
				\draw[quatcolor, very thick] (\xQc,{(0.3757-0.30)/0.70*5.71}) -- (\xQc,{(0.3827-0.30)/0.70*5.71});
				\node at (\xQc,{(0.3792-0.30)/0.70*5.71}) {\MeanQuat{quatcolor}};
				
				\draw[qnoentcolor, very thick] (\xNc,{(0.3433-0.30)/0.70*5.71}) -- (\xNc,{(0.3587-0.30)/0.70*5.71});
				\node at (\xNc,{(0.3510-0.30)/0.70*5.71}) {\MeanBloch{qnoentcolor}};
				
				\draw[qentcolor, very thick] (\xEc,{(0.3467-0.30)/0.70*5.71}) -- (\xEc,{(0.3517-0.30)/0.70*5.71});
				\node at (\xEc,{(0.3492-0.30)/0.70*5.71}) {\MeanBlochE{qentcolor}};
				
				% ---- CIFAR-10 (512-D CNN) ---- NEW!
				\draw[realcolor, very thick] (\xRd,{(0.4643-0.30)/0.70*5.71}) -- (\xRd,{(0.4783-0.30)/0.70*5.71});
				\node at (\xRd,{(0.4713-0.30)/0.70*5.71}) {\MeanReLU{realcolor}};
				
				\draw[quatcolor, very thick] (\xQd,{(0.4567-0.30)/0.70*5.71}) -- (\xQd,{(0.4603-0.30)/0.70*5.71});
				\node at (\xQd,{(0.4581-0.30)/0.70*5.71}) {\MeanQuat{quatcolor}};
				
				\draw[qnoentcolor, very thick] (\xNd,{(0.4077-0.30)/0.70*5.71}) -- (\xNd,{(0.4265-0.30)/0.70*5.71});
				\node at (\xNd,{(0.4171-0.30)/0.70*5.71}) {\MeanBloch{qnoentcolor}};
				
				\draw[qentcolor, very thick] (\xEd,{(0.2928-0.30)/0.70*5.71}) -- (\xEd,{(0.3564-0.30)/0.70*5.71});
				\node at (\xEd,{(0.3246-0.30)/0.70*5.71}) {\MeanBlochE{qentcolor}};
				
			\end{scope}
			
			% ==================================================
			% (b) Log Training Time
			% ==================================================
			\begin{scope}[yshift=-7cm]
				
				\draw[axis] (0,0) -- (\xmax,0);
				\draw[axis] (0,0) -- (0,\ymax);
				
				% Rotated y-axis label (centered)
				\node[rotate=90, anchor=south, align=center] at (-1.15,0.5*\ymax)
				{$\ln$(Training Time)\\{\scriptsize (seconds)}};
				
				% Grid + y tick labels (natural log)
				\foreach \y/\lab in {0/2.3,1.43/4.9,2.86/7.5,4.29/10.1,5.71/12.7}{
					\draw[grid] (0,\y) -- (\xmax-0.2,\y);
					\node[left, font=\scriptsize] at (0,\y) {\lab};
				}
				
				\draw[sep] (\xsep,0) -- (\xsep,\ymax);
				\draw[sep] (\xseptwo,0) -- (\xseptwo,\ymax);
				\draw[sep] (\xsepthree,0) -- (\xsepthree,\ymax);
				
				% Dataset labels
				\node[font=\bfseries] at (0.125*\xmax,\ymax+0.35) {MNIST};
				\node[font=\bfseries] at (0.375*\xmax,\ymax+0.35) {FashionMNIST};
				\node[font=\bfseries, align=center] at (0.625*\xmax,\ymax+0.35) {CIFAR-10\\{\scriptsize (16-D)}};
				\node[font=\bfseries, align=center] at (0.875*\xmax,\ymax+0.35) {CIFAR-10\\{\scriptsize (512-D)}};
				
				% Mapping: y = (ln(time) - 2.3) / 10.4 * 5.71
				
				% ---- MNIST ----
				\draw[realcolor, very thick] (\xR,{(3.08-2.3)/10.4*5.71}) -- (\xR,{(3.56-2.3)/10.4*5.71});
				\node at (\xR,{(3.40-2.3)/10.4*5.71}) {\MeanReLU{realcolor}};
				
				\draw[quatcolor, very thick] (\xQ,{(4.27-2.3)/10.4*5.71}) -- (\xQ,{(4.46-2.3)/10.4*5.71});
				\node at (\xQ,{(4.35-2.3)/10.4*5.71}) {\MeanQuat{quatcolor}};
				
				\draw[qnoentcolor, very thick] (\xN,{(9.53-2.3)/10.4*5.71}) -- (\xN,{(10.39-2.3)/10.4*5.71});
				\node at (\xN,{(10.11-2.3)/10.4*5.71}) {\MeanBloch{qnoentcolor}};
				
				\draw[qentcolor, very thick] (\xE,{(10.13-2.3)/10.4*5.71}) -- (\xE,{(10.74-2.3)/10.4*5.71});
				\node at (\xE,{(10.50-2.3)/10.4*5.71}) {\MeanBlochE{qentcolor}};
				
				% ---- FashionMNIST ----
				\draw[realcolor, very thick] (\xRr,{(3.31-2.3)/10.4*5.71}) -- (\xRr,{(3.56-2.3)/10.4*5.71});
				\node at (\xRr,{(3.47-2.3)/10.4*5.71}) {\MeanReLU{realcolor}};
				
				\draw[quatcolor, very thick] (\xQr,{(3.92-2.3)/10.4*5.71}) -- (\xQr,{(4.18-2.3)/10.4*5.71});
				\node at (\xQr,{(4.04-2.3)/10.4*5.71}) {\MeanQuat{quatcolor}};
				
				\draw[qnoentcolor, very thick] (\xNr,{(9.92-2.3)/10.4*5.71}) -- (\xNr,{(10.15-2.3)/10.4*5.71});
				\node at (\xNr,{(10.01-2.3)/10.4*5.71}) {\MeanBloch{qnoentcolor}};
				
				\draw[qentcolor, very thick] (\xEr,{(9.62-2.3)/10.4*5.71}) -- (\xEr,{(10.82-2.3)/10.4*5.71});
				\node at (\xEr,{(10.45-2.3)/10.4*5.71}) {\MeanBlochE{qentcolor}};
				
				% ---- CIFAR-10 (16-D) ----
				\draw[realcolor, very thick] (\xRc,{(2.25-2.3)/10.4*5.71}) -- (\xRc,{(2.55-2.3)/10.4*5.71});
				\node at (\xRc,{(2.40-2.3)/10.4*5.71}) {\MeanReLU{realcolor}};
				
				\draw[quatcolor, very thick] (\xQc,{(3.08-2.3)/10.4*5.71}) -- (\xQc,{(3.30-2.3)/10.4*5.71});
				\node at (\xQc,{(3.19-2.3)/10.4*5.71}) {\MeanQuat{quatcolor}};
				
				\draw[qnoentcolor, very thick] (\xNc,{(9.59-2.3)/10.4*5.71}) -- (\xNc,{(9.71-2.3)/10.4*5.71});
				\node at (\xNc,{(9.65-2.3)/10.4*5.71}) {\MeanBloch{qnoentcolor}};
				
				\draw[qentcolor, very thick] (\xEc,{(9.75-2.3)/10.4*5.71}) -- (\xEc,{(10.35-2.3)/10.4*5.71});
				\node at (\xEc,{(10.05-2.3)/10.4*5.71}) {\MeanBlochE{qentcolor}};
				
				% ---- CIFAR-10 (512-D CNN) ---- NEW!
				% ln(14.2) ≈ 2.65, ln(36.1) ≈ 3.59, ln(64739) ≈ 11.08, ln(66114) ≈ 11.10
				\draw[realcolor, very thick] (\xRd,{(2.50-2.3)/10.4*5.71}) -- (\xRd,{(2.80-2.3)/10.4*5.71});
				\node at (\xRd,{(2.65-2.3)/10.4*5.71}) {\MeanReLU{realcolor}};
				
				\draw[quatcolor, very thick] (\xQd,{(3.58-2.3)/10.4*5.71}) -- (\xQd,{(3.60-2.3)/10.4*5.71});
				\node at (\xQd,{(3.59-2.3)/10.4*5.71}) {\MeanQuat{quatcolor}};
				
				\draw[qnoentcolor, very thick] (\xNd,{(10.68-2.3)/10.4*5.71}) -- (\xNd,{(11.48-2.3)/10.4*5.71});
				\node at (\xNd,{(11.08-2.3)/10.4*5.71}) {\MeanBloch{qnoentcolor}};
				
				\draw[qentcolor, very thick] (\xEd,{(10.95-2.3)/10.4*5.71}) -- (\xEd,{(11.25-2.3)/10.4*5.71});
				\node at (\xEd,{(11.10-2.3)/10.4*5.71}) {\MeanBlochE{qentcolor}};
				
			\end{scope}
			
			% ==================================================
			% Legend (now uses the same glyph markers)
			% ==================================================
			\begin{scope}[yshift=-9.0cm, xshift=2.5cm]
				\node at (0,0) {\MeanReLU{realcolor}};
				\node[right] at (0.35,0) {RealNet};
				
				\node at (2.7,0) {\MeanQuat{quatcolor}};
				\node[right] at (3.05,0) {QuatNet};
				
				\node at (5.8,0) {\MeanBloch{qnoentcolor}};
				\node[right] at (6.15,0) {Quantum--NoEnt};
				
				\node at (9.5,0) {\MeanBlochE{qentcolor}};
				\node[right] at (9.75,0) {Quantum--Ent};
			\end{scope}
			
		\end{tikzpicture}
		
\caption{Performance across three random seeds on MNIST, FashionMNIST, and CIFAR-10 under two frozen feature regimes. Top panels show test accuracy, and bottom panels show log-scaled training time. Bars show seed ranges and markers denote means.}

		\label{fig:performance_comparison}
		
	\end{adjustwidth}
\end{figure}
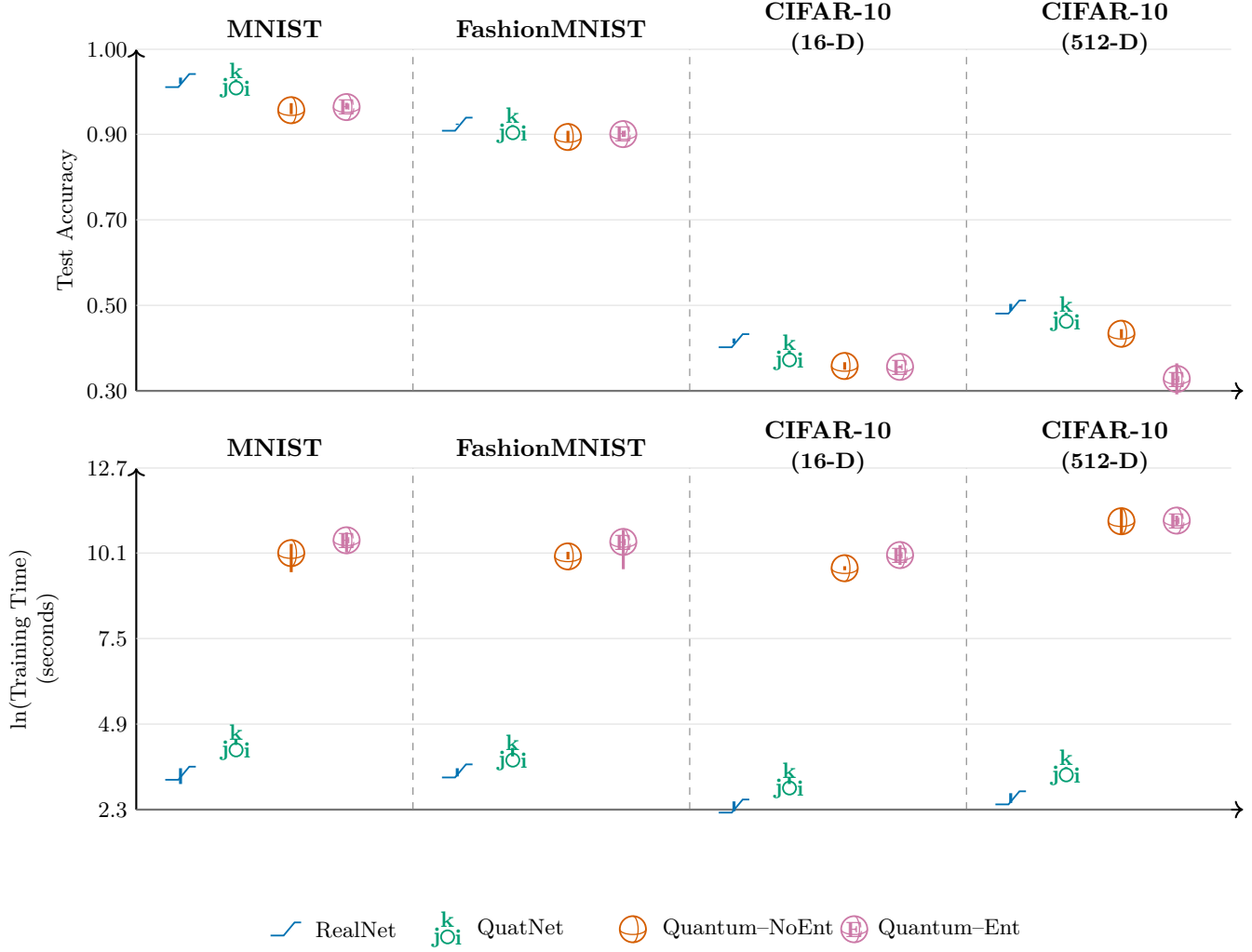

Table~\ref{tab:aggregated_results} and Fig.~\ref{fig:performance_comparison} summarize model performance across accuracy, variance, and computational cost. On MNIST and FashionMNIST, QuatNet matches RealNet within statistical variability (differences $\leq0.2$ percentage points) despite fewer parameters (1{,}000 vs.\ 1{,}738). Both quantum models underperform the classical heads, with the largest grayscale gap occurring on MNIST and smaller but consistent gaps observed on FashionMNIST.

On CIFAR-10 with the learned 16-dimensional bottleneck, all models decline in absolute accuracy due to increased task complexity, but the relative ordering persists. RealNet achieves $40.23\%\pm0.44\%$, QuatNet retains $94.26\%$ of this performance ($37.92\%\pm0.33\%$), Quantum-NoEnt reaches $35.10\%\pm0.71\%$, and Quantum-Ent achieves $34.92\%\pm0.22\%$, marginally below the product-state variant and within one standard deviation.

Using frozen 512-dimensional ResNet18 features increases classical-model accuracy while preserving the ordering. RealNet reaches $47.13\%$, QuatNet retains $97.20\%$ of this performance ($45.81\%$), Quantum-NoEnt achieves $41.71\%$, and Quantum-Ent drops to $32.46\%$, underperforming the product-state variant by $9.25$ percentage points.

Across datasets and feature regimes, the same broad pattern emerges: RealNet and QuatNet form the top-performing group, while both shallow quantum variants remain below the classical heads. The relative ordering of Quantum-NoEnt and Quantum-Ent depends on dataset and feature regime. Quantum models also exhibit substantially higher training cost and greater optimization instability. These results indicate that quaternion-valued networks, whose classification heads share local $\mathrm{SU}(2)$ parameterization with single-qubit gates, closely match real-valued baselines while consistently outperforming shallow quantum circuits on classical vision tasks.

\begin{table*}[!htbp]
	\centering
	\small
\caption{Aggregated test performance under shared frozen preprocessors, reported as mean $\pm$ standard deviation over three seeds. MNIST and FashionMNIST use a frozen 16-dimensional bottleneck, while CIFAR-10 is evaluated using both the bottleneck and frozen 512-dimensional ImageNet-pretrained ResNet18 features. Times report wall-clock training time per seed, and parameter counts exclude frozen backbone parameters.}
	\label{tab:aggregated_results}
	
	\resizebox{\textwidth}{!}{%
		\begin{tabular}{llcccc}
			\toprule
			\textbf{Dataset} & \textbf{Model}
			& \textbf{Accuracy}
			& \textbf{Epochs}
			& \textbf{Time (s)}
			& \textbf{Trainable Params} \\
			\midrule
			
			\multirow{4}{*}{MNIST}
			& RealNet        & $0.9354 \pm 0.0052$ & $21.3 \pm 4.5$ & $30.0 \pm 5.9$          & $1{,}738$ \\
			& QuatNet        & $\mathbf{0.9364 \pm 0.0015}$ & $31.7 \pm 2.4$ & $77.4 \pm 6.5$   & $1{,}000$  \\
			& Quantum-NoEnt  & $0.8752 \pm 0.0102$ & $51.3 \pm 16.4$& $24{,}601 \pm 7{,}891$  & $162$      \\
			& Quantum-Ent    & $0.8816 \pm 0.0062$ & $67.7 \pm 16.0$& $36{,}492 \pm 8{,}670$  & $162$      \\
			\midrule
			
			\multirow{4}{*}{FashionMNIST}
			& RealNet        & $\mathbf{0.8460 \pm 0.0012}$ & $23.0 \pm 2.2$ & $32.0 \pm 3.3$          & $1{,}738$ \\
			& QuatNet        & $0.8447 \pm 0.0005$ & $24.0 \pm 2.4$ & $56.9 \pm 6.3$  & $1{,}000$  \\
			& Quantum-NoEnt  & $0.8203 \pm 0.0094$ & $47.3 \pm 4.0$ & $22{,}285 \pm 2{,}414$  & $162$      \\
			& Quantum-Ent    & $0.8262 \pm 0.0052$ & $66.3 \pm 27.9$& $34{,}446 \pm 14{,}570$ & $162$      \\
			\midrule
			
			\multirow{4}{*}{\shortstack[l]{CIFAR-10\\(16-D bottleneck)}}
			& RealNet        & $\mathbf{0.4023 \pm 0.0044}$ & $24.0 \pm 3.6$ & $10.3 \pm 1.3$  & $1{,}738$ \\
			& QuatNet        & $0.3792 \pm 0.0033$ & $36.0 \pm 5.7$ & $24.3 \pm 2.7$  & $1{,}000$  \\
			& Quantum-NoEnt  & $0.3510 \pm 0.0071$$^{\dagger}$ & $32.3 \pm 1.9$ & $15{,}504 \pm 917$ & $162$  \\
			& Quantum-Ent    & $0.3492 \pm 0.0022$ & $43.3 \pm 14.5$& $23{,}117 \pm 7{,}498$ & $162$ \\
			\midrule
			
			\multirow{4}{*}{\shortstack[l]{CIFAR-10\\(ResNet18 512-D)}}
			& RealNet        & $\mathbf{0.4713 \pm 0.0070}$ & $22.3 \pm 4.0$ & $14.2 \pm 2.5$ & $66{,}954$ \\
			& QuatNet        & $0.4581 \pm 0.0016$ & $61.7 \pm 8.3$ & $53.3 \pm 7.1$ & $17{,}832$ \\
			& Quantum-NoEnt  & $0.4171 \pm 0.0094$ & $48.7 \pm 10.9$ & $64{,}739 \pm 13{,}400$ & $4{,}282$ \\
			& Quantum-Ent    & $0.3246 \pm 0.0318$ & $42.7 \pm 3.9$ & $66{,}114 \pm 5{,}306$ & $4{,}282$ \\
			
			\bottomrule
			
\multicolumn{6}{l}{{\scriptsize $^\dagger$Robustness check at 2{,}000 samples/class: accuracy $0.3541 \pm 0.0054$, within 0.3 pp of reported figure; variance modestly reduced.}}\\
			\bottomrule
		\end{tabular}%
	}
\end{table*}

\FloatBarrier

\subsection*{Robustness across seeds and sample sizes}

The five-seed MNIST extension preserves the primary performance ordering across all seeds. RealNet and QuatNet achieve $93.47\% \pm 0.18\%$ and $93.45\% \pm 0.15\%$, respectively, indicating near-equivalent performance (gap $= 0.02$ pp, $d = 0.202$). In contrast, Quantum-NoEnt achieves $87.31\% \pm 1.06\%$ and Quantum-Ent achieves $88.17\% \pm 0.79\%$, remaining substantially below both classical baselines. A Friedman test provides evidence that the MNIST five-seed model ordering is unlikely to be random ($\chi^2 = 12.796$, $p = 0.0051$, $n = 5$). Post-hoc Wilcoxon signed-rank tests yield $p = 0.0625$ for both QuatNet vs.\ Quantum-NoEnt ($d = 5.736$) and QuatNet vs.\ Quantum-Ent ($d = 6.403$). At $n = 5$, this corresponds to the minimum attainable two-sided Wilcoxon $p$-value, while the large effect sizes ($d > 5$) provide scale-independent evidence of substantial performance separation. For FashionMNIST and CIFAR-10, where only three seeds are available, effect sizes serve as the primary inferential statistic. QuatNet versus Quantum comparisons produce consistently large effects ($d > 2.0$) across both datasets, whereas $p = 0.25$ reflects the minimum attainable two-sided Wilcoxon $p$-value at $n = 3$. Near-equivalence between RealNet and QuatNet is also observed on FashionMNIST (gap $= 0.13$ pp). The RealNet versus QuatNet Cohen's $d$ for CIFAR-10 is omitted because the near-zero paired differences produce numerical instability in the standardized effect size estimate; specifically, the pooled standard deviation approaches the noise floor, rendering the ratio unstable. 

We next examine variance stability and sample-size sensitivity under frozen ResNet18 features, where Quantum-Ent exhibited the most pronounced instability in the primary analysis. On CIFAR-10 with frozen ResNet18 features, QuatNet exhibits lower seed-to-seed variability than RealNet and the quantum models in the frozen ResNet18 regime ($0.16$ percentage points for QuatNet, $0.70$ percentage points for RealNet, $0.94$ percentage points for Quantum-NoEnt, and $3.18$ percentage points for Quantum-Ent), suggesting that the unit-quaternion $\mathrm{SU}(2)$ constraint may contribute to more stable optimization under high-dimensional representations. The elevated variance of Quantum-Ent under pretrained features is notable: a standard deviation of $3.18\%$ across three seeds is four to eight times that of the classical baselines, and is consistent with the non-monotonic scaling behavior observed in the sample-size excursion at 1{,}500, 2{,}000, and 2{,}500 training samples per class (Table~\ref{tab:scaling}).

To examine whether the observed Quantum-Ent degradation under frozen ResNet18 features reflects data starvation rather than architectural constraints, we conducted two complementary robustness evaluations. First, we evaluated seed sensitivity on CIFAR-10 using the primary 1{,}500-sample configuration across three random seeds. Second, we performed a targeted sample-scaling analysis of Quantum-Ent and QuatNet at 1{,}500, 2{,}000, and 2{,}500 training samples per class under otherwise identical conditions (seed~42, patience~10, maximum 200 epochs). These evaluations deliberately perturb initialization and training set size to determine whether the performance ordering and variance patterns observed in the primary analysis remain stable under increased data exposure. Table~\ref{tab:scaling} summarizes the scenario results alongside the primary 1{,}500-sample figures from Table~\ref{tab:aggregated_results}.

QuatNet exhibited stable, near-saturated performance across all sample sizes on CIFAR-10, with accuracy ranging from $45.81\%$ to $46.54\%$ --- a span of 0.73 percentage points across a 67\% increase in training data --- and completing all three seeds in under 90 seconds at every sample size. Quantum-Ent showed markedly different behavior: accuracy rose from $32.46\% \pm 3.18\%$ at 1{,}500 samples to $39.07\%$ at 2{,}000 samples, then degraded to $38.13\%$ at 2{,}500 samples, exhibiting a non-monotonic scaling pattern that required 136{,}563 seconds (approximately 38 GPU-hours) for a single seed. New best accuracies were recorded as late as epochs 25, 35, 40, 41, and 45 during the 2{,}500-sample run, contrasting sharply with QuatNet's clean monotonic convergence.

The seed and sample scaling results provide convergent evidence that the observed performance ordering is stable under perturbation along both dimensions. QuatNet accuracy is essentially flat across seeds and sample sizes, with consistently low variance throughout. Quantum-Ent performance is non-monotonic with respect to sample size, erratic in optimization trajectory, and extraordinarily expensive to verify --- with Quantum-Ent requiring more than $1{,}600\times$ the total wall-clock time of QuatNet across the scaling study. This differential stability, observed across the evaluated dataset-feature regimes and robustness checks, is consistent with the interpretation that quantum underperformance is not explained by sampling artifacts or initialization sensitivity alone.

\begin{table*}[!htbp]
	\centering
	\small
\caption{Sample-size sensitivity under frozen ResNet18 features on CIFAR-10. QuatNet results are reported as mean $\pm$ standard deviation over three seeds at all sample sizes. Quantum-Ent results are reported as mean $\pm$ standard deviation at 1{,}500 samples per class and as single-seed estimates at 2{,}000 and 2{,}500 samples per class due to computational cost.}
	\label{tab:scaling}
	\begin{tabular}{llcc}
		\toprule
		\textbf{Model} & \textbf{Samples/class} & \textbf{Accuracy} & \textbf{Time (s)} \\
		\midrule
		QuatNet     & 1{,}500 & $0.4581 \pm 0.0016$ & $53 \pm 7$    \\
		QuatNet     & 2{,}000 & $0.4647 \pm 0.0018$ & $71 \pm 7$    \\
		QuatNet     & 2{,}500 & $0.4654 \pm 0.0008$ & $77 \pm 5$    \\
		\midrule
		Quantum-Ent & 1{,}500 & $0.3246 \pm 0.0318$ & $66{,}114$    \\
		Quantum-Ent & 2{,}000 & $0.3907$             & $92{,}168$    \\
		Quantum-Ent & 2{,}500 & $0.3813$             & $136{,}563$   \\
		\bottomrule
	\end{tabular}
\end{table*}

\FloatBarrier

We note an asymmetry in the evaluation design: QuatNet results reflect mean $\pm$ standard deviation across three seeds at all sample sizes, whereas Quantum-Ent results are single-seed at 2{,}000 and 2{,}500 samples per class, reflecting the computational cost of approximately 18--38 GPU-hours per seed in the ResNet18 regime. The non-monotonic scaling pattern and erratic optimization trajectory observed for Quantum-Ent are therefore indicative rather than statistically definitive and should be interpreted alongside the three-seed variance already documented at 1{,}500 samples ($\pm 3.18\%$) in Table~\ref{tab:aggregated_results}.

%%%%%%%% DISCUSSION %%%%%%%%%%

\section*{Discussion}

The experiments identify a regime in which classical implementations of local $\mathrm{SU}(2)$ structure match real-valued baselines, while shallow variational quantum circuits exhibit persistent optimization and measurement limitations. Most notably, entanglement provides only modest benefit on simple datasets and reverses sharply under richer pretrained representations, supporting the interpretation that shallow entangling dynamics alone are insufficient to confer practical advantage on classical vision tasks.

\subsection*{Quaternion networks match classical baselines and exceed quantum models}

Quaternion-valued classification heads closely match real-valued baselines while using substantially fewer trainable parameters. On MNIST, the five-seed evaluation ($n = 5$) yields $93.45\% \pm 0.15\%$ for QuatNet versus $93.47\% \pm 0.18\%$ for RealNet, supporting near-equivalence between the two models (gap $= 0.02$ pp). On FashionMNIST, the three-seed primary results show $84.47\%$ versus $84.60\%$, a difference of 0.13 percentage points. On CIFAR-10, QuatNet retains $94.26\%$ of real-valued performance under the learned bottleneck and $97.20\%$ under frozen ResNet18 features, indicating stable behavior across a 32-fold increase in feature dimensionality. A Friedman test on the five-seed MNIST evaluation indicates that the overall model ordering is unlikely to be random ($\chi^2 = 12.796$, $p = 0.0051$, $n = 5$). On FashionMNIST and CIFAR-10, large effect sizes ($d > 2.0$) for all QuatNet vs.\ quantum comparisons serve as the primary inferential statistic given $n = 3$.

These results are achieved with reduced head capacity: 1{,}000 versus 1{,}738 trainable parameters on MNIST/FashionMNIST and 17{,}832 versus 66{,}954 on CIFAR-10 with ResNet18 features. QuatNet also exhibits consistently low variance across seeds, particularly under pretrained features, indicating that unit-quaternion $\mathrm{SU}(2)$ constraints may stabilize optimization in high-dimensional feature spaces \cite{kuipers1999quaternions}.

The ResNet18 experimental setting compresses 512-dimensional frozen features into an eight-qubit quantum representation with 12 measured observables, creating an information bottleneck that may affect the comparison between classical and quantum models. Structure-preserving compression methods, such as PCA-informed encodings or amplitude encoding schemes that better preserve geometric relationships, may provide fairer quantum encodings of high-dimensional feature vectors. Richer measurement strategies, including generalized POVM measurements or classical shadow protocols, may also reduce information loss relative to the limited set of measured observables used here. These alternatives represent important directions for future work. The present results characterize the shallow-VQC regime as evaluated, which we consider a practically relevant baseline for near-term quantum advantage claims on classical vision data.

The performance degradation observed in the Quantum-Ent model and its non-monotonic training trajectory may reflect several interacting architectural and optimization constraints. Shallow circuit depth limits the expressivity of the variational ansatz, while ring-entanglement topology restricts the reachable set of entangled states and may introduce inter-qubit correlations that are not well aligned with the task structure. These constraints can also produce a difficult optimization landscape, which may help explain why Quantum-Ent does not improve monotonically with training progress or with increased sample size in the extended CIFAR-10 experiments. We therefore interpret the observed behavior as consistent with structural limitations of the shallow entangled architecture, rather than as stochastic training noise alone.

In contrast, shallow product-state quantum circuits underperform quaternion networks despite operating on identical frozen features. This gap suggests that shared local $\mathrm{SU}(2)$ geometry --- even as implemented in full quantum circuits with measurement and Hilbert-space evolution --- is insufficient to overcome the expressivity constraints characteristic of shallow variational circuits on classical vision tasks.

\subsection*{Entanglement: modest gains on simple features, reversal on rich features}

Introducing entanglement through CNOT operations yields consistent but modest improvements over product-state circuits on MNIST and FashionMNIST, increasing accuracy by approximately $0.6$ percentage points on both datasets. These gains are reproducible across seeds and are accompanied by reduced variance relative to Quantum-NoEnt, signaling that shallow entanglement may partially stabilize optimization in low-complexity regimes \cite{holmes2022connecting}.

On CIFAR-10 with the learned bottleneck, entanglement provides no measurable benefit. Under frozen ResNet18 features, Quantum-Ent collapses to $32.46\% \pm 3.18\%$, underperforming the product-state variant by 9.25 percentage points with substantially higher variance, and this degradation persists and remains non-monotonic as sample size increases from 1{,}500 to 2{,}500 per class --- while QuatNet remains stable at $45.81\%$--$46.54\%$ across the same range. The reversal is therefore consistent with architectural and measurement constraints rather than being explained by data availability alone.

This reversal suggests that shallow entanglement interacts adversely with rich classical representations, amplifying optimization instability and measurement bottlenecks rather than improving expressivity. Although entangling gates introduce correlations unavailable to product-state circuits, shallow depth and restricted observables appear insufficient to translate these correlations into improved learning on higher-complexity tasks.

\subsection*{Measurement bottlenecks and shallow-circuit limits}

Quaternion networks consistently outperform product-state quantum circuits despite sharing the same local $\mathrm{SU}(2)$-derived parameterization. Across all benchmarks, Quantum-NoEnt trails QuatNet by 2.4--6.1 percentage points, with substantially higher variance and orders-of-magnitude longer training times.

This gap is consistent with measurement compression and limited expressivity in shallow variational circuits serving as dominant constraints. Quaternion layers preserve four-dimensional structure through Hamilton products, whereas quantum circuits compress high-dimensional quantum states into only 6 or 12 measured observables before classification. In the product-state regime, two-qubit correlations factorize and therefore contribute no additional information beyond single-qubit measurements, further constraining effective representational capacity.

The restriction to shallow circuits is deliberate: increasing depth improves expressivity but introduces substantial optimization and scaling difficulties, including barren plateau phenomena \cite{mcclean2018barren, huembeli2021characterizing}. Within the shallow, measurement-limited regime studied here, the results suggest that measurement compression and restricted circuit expressivity limit variational quantum classifiers relative to quaternion-valued $\mathrm{SU}(2)$ models. The interpretation of performance gaps in terms of measurement compression and shallow-circuit expressivity limits, while consistent with the observed patterns, has not been experimentally isolated through ablations on observable count or circuit expressivity.

\subsection*{Optimization geometry limits variational quantum circuits}

Variational quantum circuits may face intrinsic optimization challenges arising from parameter-shift gradient estimation and sensitivity to initialization. Relative to quaternion and real-valued networks, quantum models exhibit substantially greater variance across seeds, particularly under pretrained ResNet18 features.

Controlled diagnostics (Table~\ref{tab:opt_geometry}) show that Fubini--Study / quantum Fisher information (FS/QFI) preconditioning produces update directions nearly collinear with the natural-gradient step but yields losses statistically indistinguishable from Adam at substantially higher computational cost. This suggests that curvature-aware optimization alone does not overcome the dominant optimization limitations of depth-3 circuits in the present regime.

Quaternion networks, by contrast, achieve both higher accuracy and lower variance across datasets and feature regimes, indicating that structured $\mathrm{SU}(2)$ constraints can stabilize rather than hinder optimization \cite{kuipers1999quaternions}.

\subsection*{Robustness of performance relationships across representations, sample sizes, and seeds}

Four complementary robustness evaluations --- frozen ResNet18 features on CIFAR-10, sample scaling of QuatNet and Quantum-Ent at 1{,}500--2{,}500 samples per class on CIFAR-10, a Quantum-NoEnt sample-size check at 2{,}000 samples on the learned bottleneck, and a five-seed MNIST extension --- support the same interpretation: the observed performance ordering is not explained by representation quality, sample size, or initialization alone.

QuatNet matches RealNet under every perturbation. The five-seed MNIST gap of 0.02 percentage points ($d = 0.202$) reduces concern about initialization sensitivity; stability across $45.81\%$--$46.54\%$ over a 67\% increase in training data reduces concern about sample starvation; and retention of $97.20\%$ of RealNet performance across a 32-fold increase in feature dimensionality reduces concern about representation dependence. A Friedman test on the MNIST five-seed evaluation provides evidence that the overall ordering is unlikely to be random ($\chi^2 = 12.796$, $p = 0.0051$, $n = 5$). On FashionMNIST and CIFAR-10, large effect sizes ($d > 2.0$) for all QuatNet vs.\ quantum comparisons serve as the primary inferential statistic given $n = 3$. Together these results indicate that the $\mathrm{SU}(2)$ inductive bias of quaternion layers is robust rather than regime-specific.

Quantum underperformance is equally persistent but behaviorally distinct. On MNIST, Quantum-NoEnt and Quantum-Ent remain 5--6 percentage points below both classical baselines across all five seeds, with no sign of convergence toward classical performance. On CIFAR-10, Quantum-Ent exhibits non-monotonic sample scaling --- improving from $32.46\%$ to $39.07\%$ then degrading to $38.13\%$ as samples increase from 1{,}500 to 2{,}500 --- while Quantum-NoEnt stays within 0.3 percentage points of its primary result at 2{,}000 samples. Non-monotonic scaling and erratic optimization trajectories under increased data are consistent with shallow-circuit expressivity limits and measurement compression, rather than data starvation alone.

\subsection*{Implications for quantum machine learning research}

These results suggest a stronger classical baseline for quantum machine learning on classical vision tasks. Claims of quantum advantage on comparable classical vision tasks should therefore be evaluated not only against conventional real-valued networks, but also against parameter-efficient quaternion models that share the same local $\mathrm{SU}(2)$ geometry as single-qubit operations.

The findings also narrow the plausible sources of quantum advantage on classical data. Shared local $\mathrm{SU}(2)$ geometry, even as realized in quantum circuits, is insufficient and shallow entanglement does not reliably improve performance. Entanglement provides modest gains on simple grayscale datasets, no measurable benefit on bottleneck CIFAR-10, and severe degradation under frozen pretrained features, suggesting that shallow entanglement may amplify optimization and measurement constraints rather than overcome them.

These observations align with the growing literature on quantum kernel methods, which finds that performance on classical data is often governed more by data structure than by uniquely quantum effects \cite{huang2021power,koelle2023disentangling}. Under the frozen-feature design used here, real-valued, quaternion, and quantum heads operate on identical representations and differ only in transformation geometry and parameterization. Within this regime, quaternion-valued networks consistently outperform shallow variational quantum circuits while providing a more geometrically faithful classical comparator than standard real-valued baselines.

\subsection*{Scope, limitations, and future directions}

The conclusions are bounded by several design choices. First, all benchmarks are classical vision datasets without intrinsic quantum structure. This setting is appropriate for evaluating near-term quantum advantage claims on standard supervised-learning tasks, but it does not test problems where quantum data, quantum kernels, or quantum simulation structure may provide a more natural advantage. Second, the circuits are intentionally shallow to preserve trainability. The depth-3 quantum circuits evaluated here already require substantially longer training times than quaternion classifiers while achieving lower accuracy, making systematic depth scaling infeasible within the present study. Third, the entangled architecture uses a simple CNOT ring topology. Richer entanglement schemes, hardware-efficient ansatz variants, problem-informed connectivity, or adaptive circuit designs may alter the observed tradeoffs, although they would also introduce additional optimization and computational costs.

The comparison is also limited by the encoding and measurement choices required to make the shallow-VQC benchmark computationally tractable. In the ResNet18 setting, 512-dimensional frozen features are compressed into an eight-qubit representation with 12 measured observables. This compression creates an information bottleneck that may disadvantage the quantum models relative to classical heads that preserve richer intermediate representations. Structure-preserving compression methods, amplitude encoding schemes, alternative observable sets, generalized POVM measurements, or classical-shadow-style measurement protocols may reduce this bottleneck in future work. The present results therefore characterize the shallow, measurement-limited VQC regime evaluated here rather than all possible quantum encodings or measurement strategies.

The results should also be interpreted as evidence about the tested architectures, not as a general impossibility result for quantum machine learning. Across the evaluated datasets, feature regimes, and sample-size perturbations, quaternion classifiers consistently match or approach real-valued baselines while outperforming shallow VQCs. This pattern suggests that shared local $\mathrm{SU}(2)$ geometry and shallow ring entanglement are not sufficient, by themselves, to produce practical quantum advantage on the classical vision benchmarks studied here. However, the findings do not rule out advantages from deeper circuits, different ansatz families, problem-specific encodings, larger observable sets, quantum-native datasets, or hardware implementations with properties not captured by classical simulation.

All quantum experiments use GPU-based simulation and therefore do not capture hardware noise, decoherence, gate errors, finite-shot sampling, queueing constraints, or device-specific connectivity limitations \cite{preskill2018quantum}. Simulation is appropriate for isolating architectural and optimization effects under controlled conditions, but real-hardware performance may differ. Conversely, hardware execution would introduce additional noise and sampling constraints that could either obscure or amplify the measurement and optimization limitations observed here.

Future work should extend these analyses to problems with intrinsic quantum structure, deeper and more expressive circuits, alternative measurement strategies, and real quantum hardware. Depth scaling studies should be paired with trainability methods, including quantum natural gradients \cite{stokes2020quantum}, layer-wise training, improved initialization strategies, and ansatz designs intended to mitigate barren plateau phenomena at greater depths \cite{mcclean2018barren,huembeli2021characterizing}. The instability signatures documented here, including non-monotonic sample scaling, erratic late-epoch optimization, and amplified variance under rich feature representations, provide specific behavioral patterns for such studies to test. Explicit ablations on observable count, circuit depth, entanglement topology, and encoding strategy would help isolate the relative contributions of measurement compression, shallow-circuit expressivity, and optimization difficulty. Analytical conditions identifying when classical $\mathrm{SU}(2)$ implementations are sufficient, and when genuinely quantum resources provide additional benefit, would provide a valuable theoretical complement to these empirical findings.

%%%%%%%% CONCLUSION %%%%%%%%%%
\section*{Conclusion}

This work identifies a practically relevant regime, depth-3 variational circuits on classical vision tasks with frozen feature representations, in which classical $\mathrm{SU}(2)$ models implemented through quaternion-valued neural networks match or exceed the shallow variational quantum classifiers evaluated here. Across MNIST, FashionMNIST, and CIFAR-10, quaternion networks match or closely approach real-valued MLP performance. On CIFAR-10, they retain 94--97\% of real-valued performance across the learned bottleneck and frozen ResNet18 regimes. A Friedman test on the five-seed MNIST evaluation indicates that the overall model ordering is non-random ($\chi^2 = 12.796$, $p = 0.0051$, $n = 5$), with large effect sizes ($d > 2.0$) for QuatNet versus quantum comparisons across all datasets. The same qualitative ordering is preserved across the five-seed MNIST extension, targeted sample-size perturbations, and the two feature regimes studied.

The product-state quantum classifiers evaluated here underperform quaternion networks by 2.4--6.1 percentage points, while the entangled model shows a larger degradation under frozen ResNet18 features. Across these settings, the shallow quantum classifiers require substantially greater computational time and exhibit greater optimization variability. Entanglement provides modest gains on the grayscale datasets but reverses under pretrained ResNet18 features. On CIFAR-10 with frozen ResNet18 embeddings, the entangled model performs 9.3 percentage points below the product-state circuit and exhibits non-monotonic behavior under increased data exposure. These results are consistent with the interpretation that, in this setting, shallow entanglement may amplify measurement-induced compression and optimization difficulty rather than improve task-relevant expressivity.

The observed performance gap is not explained by parameter count alone. Quaternion networks preserve continuous intermediate representations, whereas the variational quantum circuits evaluated here compress quantum states into a limited set of measured observables. Combined with shallow circuit depth and variational optimization constraints, this measurement-limited representation may restrict performance even when feature quality improves.

These conclusions are bounded to shallow, measurement-limited variational circuits simulated on classical image-classification datasets without intrinsic quantum structure. They should not be interpreted as a general limitation of quantum machine learning, deeper quantum circuits, alternative encodings, richer measurement strategies, or quantum-native learning problems. Within the evaluated regime, however, quaternion networks provide efficient and stable classical $\mathrm{SU}(2)$ baselines for assessing near-term quantum advantage claims. More broadly, the results suggest that shared local $\mathrm{SU}(2)$ geometry is not sufficient, by itself, to confer practical quantum advantage on the classical vision tasks studied here, and that the utility of entanglement depends on feature structure, circuit depth, ansatz design, and measurement strategy.

\section*{Disclosures}

\textbf{Competing Interests.} The authors have no relevant financial or non-financial interests to disclose.

\noindent
\textbf{Author Contributions.} All authors contributed equally to the conceptualization, methodology, software development, analysis, visualization, and writing of this manuscript.

\noindent
\textbf{Use of Artificial Intelligence Tools.} Generative artificial intelligence tools were used to assist with prose refinement, \LaTeX{} formatting, and figure layout. All analytical decisions, experimental design, code implementation, validation, and interpretation of results were performed by the authors, who take full responsibility for the accuracy, originality, and integrity of the work.

\noindent
\textbf{Ethics Approval and Consent to Participate.} This study does not involve human participants, human data, or animal subjects and therefore did not require ethical approval or informed consent.

\noindent
\textbf{Data and Code Availability.} All datasets used in this study are publicly available benchmark datasets. Code used to generate the results is available from the authors at \url{https://github.com/dustoff06/Quantum-Machine-Intelligence}

	%This is where your bibliography is generated. Make sure that your .bib file is actually called library.bib
	
	%\bibliographystyle{unsrt}
	\bibliography{paper}

\end{document}